\documentclass[12pt,twoside]{report}

\newcommand{\reporttitle}{CXR-Agent: Vision-language models for chest X-ray interpretation with uncertainty aware radiology reporting}

\newcommand{\reporttype}{MEng Final Year Project}
\newcommand{\degreetype}{Type of degree} 

\usepackage[a4paper,hmargin=2.8cm,vmargin=2.0cm,includeheadfoot]{geometry}
\usepackage{textpos}
\usepackage[numbers]{natbib}  
\usepackage{tabularx,longtable,multirow,subfigure,caption}
\usepackage{fancyhdr} 
\usepackage{url} 
\usepackage[english]{babel}
\usepackage{amsmath}
\usepackage{graphicx}
\usepackage{dsfont}
\usepackage{epstopdf} 
\usepackage{backref} 
\usepackage{array}
\usepackage{latexsym}
\usepackage[pdftex,pagebackref,hypertexnames=false,colorlinks]{hyperref} 
\hypersetup{pdftitle={},
  pdfsubject={}, 
  pdfauthor={},
  pdfkeywords={}, 
  pdfstartview=FitH,
  pdfpagemode={UseOutlines},
  bookmarksnumbered=true, bookmarksopen=true, colorlinks,
    citecolor=black,%
    filecolor=black,%
    linkcolor=black,%
    urlcolor=black}

\usepackage[all]{hypcap}
\usepackage[table,xcdraw]{xcolor}
\usepackage{tikz}
\usepackage{longtable}
\usepackage{booktabs}

\usepackage{float}
\usepackage{adjustbox}

\def\@makechapterhead#1{%
  \vspace*{10\p@}%
  {\parindent \z@ \raggedright \sffamily
    \interlinepenalty\@M
    \Huge\bfseries \thechapter \space\space #1\par\nobreak
    \vskip 30\p@
  }}

\def\@makechapterhead#1{%
  \vspace*{10\p@}%
  {\parindent \z@ \raggedright \sffamily
    \interlinepenalty\@M
    \Huge\bfseries \thechapter \space\space #1\par\nobreak
    \vskip 30\p@
  }}

\def\@makeschapterhead#1{%
  \vspace*{10\p@}%
  {\parindent \z@ \raggedright
    \sffamily
    \interlinepenalty\@M
    \Huge \bfseries  #1\par\nobreak
    \vskip 30\p@
  }}	
\allowdisplaybreaks

\begin{document}
\begin{titlepage}

\newcommand{\HRule}{\rule{\linewidth}{0.5mm}} 


\includegraphics[width = 4cm]{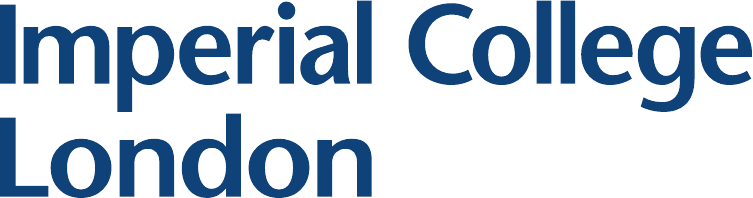}\\[0.5cm] 

\center 
 

\textsc{\LARGE \reporttype}\\[1.5cm] 
\textsc{\Large Department of Computing}\\[0.5cm] 
\textsc{\large Imperial College of Science, Technology and Medicine}\\[0.5cm] 


\HRule \\[0.4cm]
{ \huge \bfseries \reporttitle}\\ 
\HRule \\[1.5cm]
 

\begin{minipage}{0.4\textwidth}
\begin{flushleft} \large
\emph{Author:}\\
Naman Sharma
\end{flushleft}
\end{minipage}
~
\begin{minipage}{0.4\textwidth}
\begin{flushright} \large
\emph{Supervisor:} \\
Professor Ben Glocker
\end{flushright}
\end{minipage}\\[4cm]



{\large \today} 

\vfill 
Submitted in partial fulfillment of the requirements for the \degreetype~of Imperial College London

\end{titlepage}

\pagenumbering{roman}
\clearpage{\pagestyle{empty}\cleardoublepage}
\setcounter{page}{1}
\pagestyle{fancy}

\begin{abstract}

The UK's ageing population has placed extra stress on the NHS with a significant increase in chest X-rays being ordered, resulting in growing backlogs of scans pending reports. Recently large vision-language models have shown untapped potential when interpreting complex images and generating natural language descriptions using advanced cognitive reasoning. Medicine's inherently multimodal nature incorporating scans and text-based medical histories to write reports makes it conducive to benefit from these leaps in AI capabilities.\\

This project evaluates the publicly available, state of the art, foundational vision-language models for chest X-ray interpretation across several datasets and benchmarks. We use linear probes to evaluate the performance of various components including CheXagent's vision transformer and Q-former, both of which outperform the industry-standard Torch X-ray Vision models across many different datasets showing robust generalisation capabilities.  Importantly, we find that vision-language models often hallucinate with confident language, which slows down clinical interpretation. \\

Based on these findings, we develop an agent-based vision-language approach for report generation using CheXagent's linear probes and BioViL-T's phrase grounding tools to prompt a language model to generate uncertainty-aware radiology reports with pathologies localised and described based on their likelihood.\\

We thoroughly evaluate our vision-language agents using NLP metrics, chest X-ray benchmarks and clinical evaluations by developing an evaluation platform to perform a user study with respiratory specialists. Our results show considerable improvements in accuracy, interepretability and the safety of AI-generated reports. We stress the importance in analysing results for normal and abnormal scans seperately. Finally, we emphasise the need for larger paired (scan and report) datasets alongside data augmentation to tackle overfitting seen in these large vision-language models.
\end{abstract}

\clearpage
\section*{Acknowledgments}
I would like to thank Prof. Ben Glocker, Dr Fabio De Sousa Ribeiro, Dr Tian Xia and Rajat Rasal for their insight and support during the supervision of this project. Thank you to the wider BioMedIA group for creating a welcoming research environment.\\

Thank you to Dr Dominic Marshall for advising on the clinical direction for this research and evaluating the reports generated by my models. Your patience and support has helped me understand the problems that need solving. Thank you to Dr Yueqi Ge for also helping with the clinical evaluations.\\ 

Finally, thank you to my friends and family for supporting me throughout this degree. In particular, thank you to my mum, dad and sister for everything.

\clearpage{\pagestyle{empty}\cleardoublepage}

\fancyhead[RE,LO]{\sffamily {Table of Contents}}
\tableofcontents 

\clearpage{\pagestyle{empty}\cleardoublepage}
\pagenumbering{arabic}
\setcounter{page}{1}
\fancyhead[LE,RO]{\slshape \rightmark}
\fancyhead[LO,RE]{\slshape \leftmark}
\chapter{Introduction}

In the UK, the NHS is under significant stress due to an ageing population \cite{ons2024ukpopulation} with more health complications, resulting in a major increase in the number of investigations being ordered by doctors. AI automation has the scope to streamline workflows, tackle waiting lists, and reduce the cognitive load on practitioners. Large multi-modal models, have demonstrated advanced cognitive reasoning and n-shot multi-modal learning. Medicine's inherently multimodal nature incorporating scans and text-based medical histories makes it conducive to benefit from these recent leaps in AI capabilities. In particular, I explore applications of vision-language models in chest X-ray reporting workflows.\\

\section{Objectives}
\label{Objectives}
We outline the objectives we set out to achieve:
\begin{itemize}
 \item Understand and evaluate the state of the art (SOTA) large vision-language models (VLMs) for chest X-ray (CXR) interpretation. 
 \item Work with clinical experts to understand barriers to entry or shortcomings of these VLMs
 \item Tackle shortcomings in the SOTA for static CXR interpretation  (i.e. scans with no prior scan for comparison) within data and compute constraints, focusing on improvements for clinical interpretability

\end{itemize}

\section{Novel Contributions}
Our novel contributions include:
\begin{itemize}
 \item Uncertainty aware radiology reporting to improve clinical interpretability of generated reports
 \item Using linear probes to thoroughly analyse a large vision-language model (CheXagent) on various downstream tasks to find performance bottlenecks across the architecture
 \item Use of a vision encoder from a foundational model as a tool in agent-based radiology reporting workflows
 \item Analysis of various language models from domain-specific to large general models for report generation in medical agent workflows
 \item Using a vision-language model supporting phrase grounding (BioViL-T) as a pathology localisation tool in medical agent workflows
\end{itemize}

\section{Ethical Issues}

\def\checkmark{\tikz\fill[scale=0.4](0,.35) -- (.25,0) -- (1,.7) -- (.25,.15) -- cycle;} 

\begin{longtable}[!ht]{|p{0.8\linewidth}|l|l|}
    \hline
        ~ & Yes & No \\ \hline
        \cellcolor[HTML]{EFEFEF}Section 1: HUMANS & ~ & ~ \\ \hline
        Does your project involve human participants? & ~ & \checkmark \\ \hline
        \cellcolor[HTML]{EFEFEF}Section 2: PROTECTION OF PERSONAL DATA & ~ & ~  \\ \hline
        Does your project involve personal data collection and/or processing? & \checkmark & ~ \\ \hline
        Does it involve the collection and/or processing of sensitive personal data (e.g. health, sexual lifestyle, ethnicity, political opinion, religious or philosophical conviction)? & \checkmark & ~ \\ \hline
        Does it involve processing of genetic information? & ~ & \checkmark \\ \hline
        Does it involve tracking or observation of participants? & ~ & \checkmark \\ \hline
        Does your project involve further processing of previously collected personal data(secondary use)? For example Does your project involve merging existing data sets? & \checkmark & ~ \\ \hline
        \cellcolor[HTML]{EFEFEF}Section 3: ANIMALS & ~ & ~ \\ \hline
        Does your project involve animals? & ~ & \checkmark \\ \hline
        \cellcolor[HTML]{EFEFEF}Section 4: DEVELOPING COUNTRIES & ~ & ~\\ \hline
        Does your project involve developing countries? & ~ & \checkmark \\ \hline
        If your project involves low and/or lower-middle income countries, are any benefit-sharing actions planned? & ~ & \checkmark \\ \hline
        Could the situation in the country put the individuals taking part in the project at risk? & ~ & \checkmark\\ \hline
        \cellcolor[HTML]{EFEFEF}Section 5: ENVIRONMENTAL PROTECTION AND SAFETY & ~ & ~\\ \hline
        Does your project involve the use of elements that may cause harm to the environment, animals or plants? &  &  \checkmark \\ \hline
        Does your project involve the use of elements that may cause harm to humans, including project staff? & ~ & \checkmark \\ \hline
        \cellcolor[HTML]{EFEFEF}Section 6: DUAL USE & ~ & ~\\ \hline
        Does your project have the potential for military applications? & ~ & \checkmark  \\ \hline
        Does your project have an exclusive civilian application focus? & \checkmark & ~ \\ \hline
        Will your project use or produce goods or information that will require export licenses in accordance with legislation on dual use items? & ~ & \checkmark \\ \hline
        Does your project affect current standards in military ethics – e.g., global ban on weapons of mass destruction, issues of proportionality, discrimination of combatants and accountability in drone and autonomous robotics developments, incendiary or laser weapons? & ~ & \checkmark \\ \hline
        \cellcolor[HTML]{EFEFEF}Section 7: MISUSE & ~ & ~ \\ \hline
        Does your project have the potential for malevolent/criminal/terrorist abuse? & ~ & \checkmark \\ \hline
        Does your project involve information on/or the use of biological-, chemical-, nuclear/radiological-security sensitive materials and explosives, and means of their delivery? & ~ & \checkmark \\ \hline
        Does your project involve the development of technologies or the creation of information that could have severe negative impacts on human rights standards (e.g. privacy, stigmatization, discrimination), if misapplied? & ~ & \checkmark \\ \hline
        Does your project have the potential for terrorist or criminal abuse e.g. infrastructural vulnerability studies, cybersecurity related project? & ~ & \checkmark \\ \hline
        \cellcolor[HTML]{EFEFEF}SECTION 8: LEGAL ISSUES & ~ & ~ \\ \hline
        Will your project use or produce software for which there are copyright licensing implications? & ~ & \checkmark \\ \hline
        Will your project use or produce goods or information for which there are data protection, or other legal implications? &\checkmark & ~ \\ \hline
        \cellcolor[HTML]{EFEFEF}SECTION 9: OTHER ETHICS ISSUES & ~ & ~\\ \hline
        Are there any other ethics issues that should be taken into consideration? & ~ & \checkmark \\ \hline
        \caption{Ethical Issue Checklist from Imperial College London Department of Computing}
\end{longtable}


\chapter{Background}

This chapter provides an overview of the medical context for chest X-rays and foundational machine learning principles for vision-language models, focusing on computer vision, natural language processing and multi-modal representation learning.

\section{Medical context}
The medical field encompasses the process of diagnosing diseases and conditions by analyzing patient history, evaluating symptoms, interpreting medical scans, and considering any other pertinent data available to healthcare professionals. Diagnosis in medicine involves a comprehensive assessment, combining clinical expertise with the latest advancements in medical technology to ensure accurate and effective treatment plans for patients.

\subsection{Chest X-rays (CXR) and radiologist reports}
X-rays are generally the most common imaging tests taken. Every year, 2  billion chest X-rays are performed worldwide \cite{globalCXRcount}. In the UK alone, the NHS takes more than 20 million scans, with 21.9 million X-rays taken between April 2022 and March 2023, more than double the next most frequent imaging test, ultrasound, at 10.2 million in the same period\cite{nhs2023}. In particular, Chest X-rays (CXRs) were the test most commonly requested by GPs, with the number of requests increasing by 15.4 \% on the previous reporting period \cite{nhs2023}. \\

CXRs are taken for different reasons depending on the patient's level of care. Primary care, involving GPs, use CXRs for diagnosis, to rule out anything sinister, and for longer-term monitoring of chronic respiratory disorders\cite{drsharma2023interview}. Hence each patient usually gets a single X-ray taken or X-rays every 6-24 months depending on their condition and their history of progression. One-off CXRs taken for diagnosis are referred to as \textbf{\textit{static}}, whereas CXRs taken for continual short-term or long-term monitoring are referred to as \textbf{\textit{longitudinal}}.\\

On the other hand, in secondary care, CXRs are taken more frequently and for a broader range of applications because of the more complex patient cases typically seen in a hospital setting. CXR frequency can be hourly in intensive care and daily if monitoring pre/post-surgery. In this secondary care setting, medical professionals have far more data points in conjunction with CXRs to aid in patient treatment.\\ 

The CXR workflow involves a request for a scan with the requesting doctor including an "Indication" or a reason for why the scan was requested. The patient is scanned by a technician, most commonly in the Posterior-Anterior or Anterior-Posterior views with the scan and its corresponding "Indication" sent to a radiologist. The radiologist uses the "Indication" and any previous scans of the patient to write up "Findings", which is a description of the normal and abnormal observations about the most recent scan. Finally, an "Impression" is written as a conclusion providing clinical guidance by answering the indication and suggesting next steps.

\subsection{CXR datasets}

To facilitate Machine Learning and other big data analysis, datasets are being curated of CXR images with their associated radiology reports. CXR images have a specialised format known as DICOM (Digital Imaging and Communications in Medicine), which retains detailed metadata and allows for interoperability in healthcare systems, these DICOM images are generally preferred over JPEG/PNG images as JPEG contain compression artefacts and lose detail when undergoing grayscale quantisation. \\

There are several CXR datasets with the MIMIC-CXR database being the gold standard due to its size \cite{MIMIC-CXR}, containing 377,110 images from 227,835 radiographic studies (each study comes with a report) performed at the Beth Israel Deaconess Medical Center in Boston, MA between 2011 and 2016. However, a notable caveat is that these studies were all taken from patients in intensive care, not representative of all types of CXRs, for example, those taken at a primary care level for diagnosis or longer-term monitoring. CheXpert \cite{irvin2019chexpert} is another large CXR dataset however it does not have associated radiology reports instead only 14 pathology labels per scan. VinDr \cite{VinDrNguyen2022} is a smaller CXR dataset with no radiology reports however it comes with bounding boxes per pathology which is valuable for pathology localisation. However, the 15,000 training examples come with three different radiologist annotations for bounding boxes with limited consistency in pathology classifications and their bounding boxes, unlike with the test set of 3,000 images which has bounding boxes and pathologies unanimously agreed upon. \\

\begin{longtable}{|>{\centering\arraybackslash}m{3cm}|>{\centering\arraybackslash}m{3cm}|>{\centering\arraybackslash}m{3cm}|>{\centering\arraybackslash}m{3cm}|}
\hline 
& \textbf{MIMIC-CXR} & \textbf{CheXpert} & \textbf{VinDr} \\ \hline
\textbf{Year} & 2017 & 2019 & 2022 \\ \hline
\textbf{Medical Setting} & Mostly ICU patients & Inpatients and Outpatients & Inpatients\\ \hline
\textbf{Dataset Size} & 377,110 images with 227,835 reports & 224,831 images & 18,000 images \\ \hline
\textbf{Pros} & Large dataset, free-text radiology reports & Large dataset, standardized labels& Pathology bounding boxes provided \\ \hline
\textbf{Cons} & 76\% of reports have longitudinal terminology \cite{mimic-prior-scans} & No associated radiology reports released & Smaller dataset compared to MIMIC-CXR and CheXpert \\ \hline
\end{longtable}

\section{Convolutional Neural Networks (CNNs)}
Convolutional Neural Networks (CNNs) are a specialized subset of artificial neural networks designed to process data with a grid-like topology, such as images. In computer vision, CNNs are preferred for their ability to automatically and adaptively learn spatial hierarchies of features from input images. \\

Mathematically, a CNN consists of a series of convolutional layers, each applying a convolution operation to the input using a set of learnable filters (kernels). These filters are small spatially (e.g. 3x3 or 5x5), but extend through the full input depth. A 2D convolutional kernel used by CNNs with width `M`, height `N` and bias `b` can be defined as:
\begin{align*}
          (\mathbf{Image} * \mathbf{Kernel})[i, j] &= \sum_{m=0}^{M-1} \sum_{n=0}^{N-1} Image[i-m, j-n] \cdot Kernel [m, n] + b
\end{align*}

A key advantage of CNNs is weight sharing - each kernel is used at every position in the input, giving translation equivariance. This allows the network to detect meaningful features regardless of their precise location. The output of each CNN layer is a feature map summarising the presence of detected features in the input. After each convolutional layer, CNNs employs pooling layers to downsample the representations and reduce computational requirements. Additional convolutional layers then build up a hierarchical set of features. \\

Unlike fully connected networks that suffer from the curse of dimensionality and overfitting when dealing with high-dimensional image data, CNNs exploit spatial locality by enforcing a local connectivity pattern between neurons of adjacent layers.\\

In medical imaging, CNNs are often trained to identify patterns associated with specific medical conditions. They might be used to analyze X-rays, MRIs, CT scans, or pathology slides, learning from vast amounts of labelled image data where the diagnosis is known. Using these datasets, CNNs can learn to recognize various pathologies, such as tumours, fractures, or abnormalities in organ structures, by identifying the distinguishing features of each condition. \\

\subsection{ResNets}
ResNets \cite{he2015deep} addresses the problem of training very deep neural networks by introducing "skip connections" that allow gradients to flow through the network more effectively during training. These connections enable the training of networks that are much deeper than was previously possible, resulting in improved performance for a variety of computer vision tasks.\\ 

ResNets are particularly useful because they can learn from the complex, high-dimensional data generated by medical scans, such as CT, MRI, and X-ray images while mitigating the vanishing gradient problem that affects deep networks. This allows ResNets to capture subtle features that may be critical for diagnosing diseases. For instance, a variant of ResNet architecture called ResNeXt has been leveraged to detect thoracic diseases from chest X-ray images, showing promising results in the differentiation of multiple pathologies \cite{rajpurkar2017chexnet}.

\section{Transformers}
Transformers introduced in the seminal paper "Attention is all you need"\cite{Vaswani2017AttentionIA}, in particular the self-attention mechanism, have been the bedrock for recent advances in AI bringing about an era of generative AI. Whilst primary use cases have been text-based chat interfaces to Large Language Models (LLMs), there have been extensions to multiple modalities including imaging and audio. Below we discuss RNNs, which motivated the self-attention mechanism, and extensions of the self-attention mechanism to the imaging modality via the Vision Transformer.

\subsection{Limitations of RNNs}
Recurrent Neural Networks (RNNs) are a class of artificial neural networks designed to recognize patterns in sequences of data, such as text, genomes, handwriting, or time series data. Unlike traditional feedforward neural networks, RNNs contain loops allowing information to persist. This allows them to take into account the sequential nature of the input data. However, there are critical limitations to RNNs:
\begin{itemize}
    \item \textbf{Vanishing/Exploding Gradients}: Due to their recurrent nature, RNNs often suffer from vanishing or exploding gradients during training, making it hard to learn long-range dependencies within the sequence.
    \item \textbf{Sequential Computation}: The inherent sequential dependency of RNNs prevents parallelization within instances of a sequence, which leads to longer training times.
    \item \textbf{Limited Context}: Standard RNNs can find it challenging to make use of older information in the input sequence, which limits their context understanding. Additionally, RNNs mainly use the context of seen input and not future input, for example in the sentence "the boy in red shorts played football", the representation of the word "boy" will not be dependent on (or is not a function of) "football".

\end{itemize}

Transformers were introduced as a solution to the limitations of RNNs. They are a type of neural network architecture that, unlike RNNs, do not require sequential data to be processed in order. The Transformer model facilitates much higher levels of parallelization, is more efficient at capturing long-range dependencies, and generally scales better with sequence length than RNNs.

\subsection{Self-attention mechanism}
The Transformer architecture relies heavily on self-attention blocks. Self-attention is a component that updates the representation of each word or "token" in the input sequence by looking at all other tokens in the input sequence, before and after the token being updated or "queried". In other words, each token's representation becomes a function of all tokens in the input sequence, the effect of other "key" tokens on the "query" token depends on how relevant they are to the "query" token. To clarify, tokens are not words, they vary depending on the model but can be individual characters, subparts of words (OpenAI have tokens equivalent to 4 characters \cite{tokenCountOpenAI}) or multiple words, although it is most common to be shorter than a word. The reason for this variation in length is that it allows Transformers to understand words that are not in their vocabulary by understanding subwords or tokens that are. \\

The self-attention mechanism takes in a sequence of vectors, 1 vector per token, each vector is an embedding of a token and produces a sequence of vectors with new embeddings per vector. It does this by first calculating 3 new vectors per input vector, namely a query (q), key (k) and value vector (k). These 3 vectors are derived through matrix multiplication with learnable weight vectors $w_q, w_k, w_k$. To facilitate parallel computation of these updated representations, we can consider matrices of these vectors and weights (Q, K, V,  $W_q, W_k, W_k$ respectively).
\begin{align}
Q = XW_Q, \quad K = XW_K, \quad V = XW_V
\end{align}
\indent{ \textit{Note that K, Q, and V do not have to be of the same number of tokens (i.e. n dimension) as we will see later when we combine a query from an output sequence with a key and value from the input sequence.}}

Once we have computed all 3 of these matrices, we can compute the attention scores or the contextual effect one token has on another token. We consider the token being affected as the query token and the token effecting as the key token. For each pair of key and query token, we calculate its attention score as below, where d is a scaling factor equivalent to the dimensionality of the key vector to help stabilise its gradient during training:
\begin{align*}
\text{Attention Score} = \frac{q \cdot k^T}{\sqrt{d_k}} \quad
\end{align*}

If we were to use the Query and Key matrices we have, we can consider the attention score on the j\textsuperscript{th} element from the i\textsuperscript{th} element (see left diagram below). And more generally to compute all these attention scores in parallel, we can use standard matrix multiplication (see right diagram below).
\begin{align*}
\text{Attention Score}_{i,j} = \frac{Q_i \cdot K_j^T}{\sqrt{d_k}} \quad \text{Attention Score} = \frac{Q \cdot K^T}{\sqrt{d_k}} 
\end{align*}

Finally, we normalise the attention scores by passing them through a softmax layer and calculating a weighted sum for the updated final output vectors.

\begin{align*}
    \text{Attention Block Output} = \text{softmax}(\text{attention score}) \cdot V 
\end{align*}

\subsubsection{Multi-head attention}
The above attention block has one set of weights to capture a specific understanding. For example, in the sentence "Charlie loves pizza, but he hates calzone", one set of learnable weights (or one head) will focus on the positive sentiment so for Charlie it will pay more attention to loving pizza but one head cannot simultaneously pay attention to the negative sentiment (since attention is a finite sum of attention weights = 1 (due to softmax). Hence we add in another head to learn negative sentiment and so on with each head learning something new varying from local to global relationships. 
\begin{align*}
     Q^h = XW_Q^h, \quad K^h = XW_K^h, \quad V^h = XW_V^h \quad \text{for}\ h = 1 .. \text{ no. of heads}
\end{align*}
\begin{align*}
O^h = \text{Attention Weights}^h \cdot V^h 
\end{align*}
Finally, we combine the outputs of these heads to get overall multi-head attention.
\begin{align*}
\text{MultiHead}(Q, K, V ) = \text{Concat}(O^1, ..., O^H)W^O    
\end{align*}

\subsection{Transformer Architecture}

The Transformer architecture as described in "Attention is all you need" \cite{Vaswani2017AttentionIA} starts by embedding each input element. These embeddings are learned during training and are crucial to capturing the semantic meanings of input elements. Since the Transformer does not inherently capture sequence order (due to the absence of recurrence and convolution), positional encodings are added to the input embeddings. \\

\subsubsection{Encoder}
These embeddings are passed through several encoder blocks each consisting of a multi-head self-attention layer, followed by a simple feedforward network layer to help the model learn non-linear relationships within the sequence. Each sub-layer in an encoder block has a skip connection around it, followed by layer normalization. The inputs of each sub-layer are directly connected to its outputs, which helps in training deep networks by allowing gradients to flow through the network without being attenuated. These skip connections combat the vanishing gradient problem and allow for training significantly deeper models. Normalization techniques are crucial for stable and faster training. The layer normalization is applied following each skip connection, normalizing the input across the features for each data sample. This helps maintain consistent scale and variance, which is critical for effective learning in deep networks.

\subsubsection{Decoder}
The Transformer decoder block is similar to the encoder but starts with an additional sub-layer that performs masked multi-head attention over the encoder's already generated output. A casual mask is appled to prevent attention over future tokens, as these are not available at inference time.  The masking ensures predictions are dependent only on the known outputs at positions before it by masking future positions. This sub-layer's output is passed as the Query matrix into another multi-head attention which uses the encoder's output as the Value and Key matrices. The decoder returns probabilities over the token vocabulary to determine the next token.

\subsection{Vision Transformers (ViTs)}
Unlike sequences of discrete tokens, images are high-dimensional continuous values lacking an inherent notion of order. The core self-attention mechanism is preserved while tweaking embedding approaches and architecture depth for 2D images. To handle this, vision transformers split images into non-overlapping patches which are embedded and fed into the transformer analogous to tokens. Extra learnable positional embeddings are added to retain positional information tailored to the 2D image structure. Classification tokens are often appended to the sequence of image patch embeddings to provide a target for predicting image labels.

\section{Large Language Models (LLMs)}
Large language models (LLMs) are massive neural networks with billions of parameters, implemented as transformer-based architectures, and trained on huge corpora of text data in a self-supervised fashion. They are usually pretrained to predict the next word or text segment, given the previous context in an autoregressive generative language modelling objective, without the need for labelled data.

\subsection{BERT}
Bidirectional Encoder Representations from Transformers (BERT) \cite{Devlin2019BERTPO} is a transformer-based LLM designed to learn rich contextual word representations or embeddings; BERT has no decoder or output layer for text generation. BERT was trained on a vast corpus of text including the entire English Wikipedia and the BooksCorpus. It was pre-trained using two unsupervised prediction tasks: masked language modelling (where random tokens in a sentence are masked and the model learns to predict them) and next-sentence prediction. Its pre-training is resource-intensive, requiring significant computational power and managing huge text corpora.\\ 

BERT is highly versatile, capable of transferring learning to diverse downstream tasks with minimal task-specific modifications. However, BERT may struggle with very long texts due to its maximum sequence length limitation (typically 512 tokens). 

\section{Vision Language Models}
Vision language models (VLMs) are a class of large neural networks designed to understand and generate language in connection to visual concepts, unlike Diffusion models which generate images or videos. These models have become popular as they perform tasks that require a joint understanding of images and text, such as automatic image captioning, visual question answering, and generating radiology reports from medical images.
Key components include:
\begin{itemize}
    \item \textbf{Text Encoder}: To process textual information, VLMs employ language models often based on transformer architectures.
    \item \textbf{Vision Encoder}: A VLM typically uses Convolutional Neural Networks (CNNs) or Vision Transformers (ViT) to encode visual data into feature representations.
    \item \textbf{Cross-modality components}: To relate image and textual representations, a specialised co-attention mechanism is needed (in co-attention, the query is one modality and the keys/values are another). This allows the model to focus on certain parts of an image when generating or processing language, creating a joint vision-language representation in the latent space.
    \item \textbf{Decoder (Optional)}: To generate text, a VLM needs a decoder which takes the joint vision-language representation and performs auto-regressive generation of text.
\end{itemize}

\subsection{Training workflow for VLMs}
Generally, the vision and text components are pretrained separately before being integrated. Vision models are being trained on image classification/object detection tasks and text models on language modelling tasks. Then the combined VLM is pre-trained on large datasets containing pairs of images and text, for example, annotated images with descriptions.  Pre-training objectives often include masked language modelling and contrastive learning tasks, where the model learns to align the latent representations of matching images and texts while distinguishing non-matching pairs.\\

After pre-training, VLMs are fine-tuned on a specific task, such as image captioning or visual question answering, over a domain-specific dataset. During fine-tuning, the parameters of the pre-trained model are adjusted to perform optimally on the dataset.\\

\subsection{VLMs for agent workflows}
Agent workflows involve using language models to synthesize information from various sources (knowledge bases, APIs, tools, etc.) and complete tasks in a step-by-step manner. These workflows often require the language model to understand and reason over multimodal data - combining information from text, images, tables, and other formats. VLMs are well-suited for integrating visual understanding into these agent workflows. \\

By encoding the semantic information from images into a rich latent representation, VLMs can extract insights that can drive downstream reasoning and task completion. For example, in medical imaging, a VLM could analyze an X-ray and detect abnormalities. It could then feed this visual understanding to a language model agent tasked with answering questions or making diagnoses about the patient's condition. The agent could synthesize the VLM's image analysis with other data sources like the patient's medical history, lab test results, or clinical research publications (i.e. Radiopaedia \cite{radiopaedia}). This multimodal capability allows VLM-powered agents to tackle complex queries that require fusing knowledge from diverse modalities in intelligent ways.
\chapter{SOTA}
\label{SOTA}

\textbf{The SOTA section is based on research released before May 2024, with research chosen based on the availability and functionality of code. The research on foundational models (FMs) in CXR interpretation is evolving quickly, however not all work comes with code.}

\section{Key Literature}
Given the recent advancements in LMMs and the growing mismatch in supply and demand between CXR scans and radiologists to write reports, there has been a significant amount of research going into CXR reporting, from the likes of technology giants such as Google and Microsoft to smaller, independent research groups. Below we discuss how state-of-the-art developments are being made in CXR reporting, from simpler fine-tuning of pre-trained models to building and training custom architectures to handle temporal semantics inherent in radiology reporting.

\section{Fine-tuning BERT for the Medical CXR Domain} \label{CXR-BERT}
LLMs such as BERT can be pre-trained on domain-specific corpora to generate far better results for domain-specific use cases. For instance, there have been versions of the BERT architecture trained, from scratch, solely on PubMed extracts \cite{pubmedbert}, which scored 57.71\% \cite{CXRBert} against the RadNLI accuracy baseline of 53.30\% \cite{miura-etal-2021-improving} on the RadNLI inference dataset \cite{miura2021radnli}. However, to achieve SOTA performance on radiology and CXR-specific use-cases, Microsoft trained CXR-BERT \cite{CXRBert} from scratch using a corpora of PubMed extracts, MIMIC-III clinical notes and MIMIC-CXR radiology reports.\\

They first set up an adjusted vocabulary, consisting of 30,000 tokens from PubMed, MIMIC-III  and MIMIC-CXR data,  rather than using the standard BERT vocabulary as this meant medical words such as "atelectasis" would not be split into sub-tokens, such as "ate-le-ct-asis" generated by PubMedBert. This is beneficial since medical terms have specific meanings, breaking them into sub-tokens can lose semantic coherence since the fragments may not carry the same significance as the whole term. Also, breaking into smaller tokens means those sub-tokens may appear more frequently since the prefixes and suffixes in medicine are used more frequently than the complete word meaning that the model may not learn the frequency aspect of certain keywords, which is important since some conditions are far more rare in medicine. As Dr Sharma said "Everything common is common, everything rare is rare" \cite{drsharma2023interview}, meaning its important to realise that you will not see rare conditions regularly and your model should factor this in. \\

The randomly initialised BERT model is trained using a Masked Langauge modelling (MLM) objective, with dynamic whole-word masking and packing of multiple sentences into a single input sequence. This training results in the CXR-BERT-general model. This model is further specialised by continuing to pretrain solely on the MIMIC-CXR reports to create CXR-BERT-specialised, both models available on HuggingFace. These models perform far better than the PubMedBERT with a 60.46\% and 65.21\% accuracy respectively on the RadNLI inference dataset \cite{miura2021radnli}.

\section{Fine-tuning Vision Encoders for CXRs}
This task can be achieved through a variety of pre-training objectives. However, the most common ones include image-text contrastive learning and image captioning as mentioned in the \nameref{BLIP-2} section below and implemented in early 2024 by the Stanford team in their CheXagent foundational model \cite{chen2024CheXagent}.\\ 

More recently alternate methods to fine-tune ViT have been proposed by the Microsoft team in Rad-Dino's DinoV2 \cite{pérezgarcía2024raddino}. DINOv2 is a state-of-the-art self-supervised learning method for pre-training vision transformers (ViTs). It uses a teacher-student framework with both image-level and patch-level objectives to learn useful global and local representations without requiring text labels. The patch-level objective uses masked image modeling to predict masked patches. The image-level objective uses a contrastive loss over multiple crops of an image to align local features. The teacher is updated via exponential moving average of student weights. Using both local and global objectives allows DINOv2 to achieve strong performance on downstream dense prediction tasks like segmentation. Asymmetric design choices and regularization techniques like KoLeo promote robust and uniform feature learning. Fine-tuning DINOv2 for medical imaging can provide a powerful pretrained encoder for various vision tasks. The self-supervised pretraining allows learning useful representations from abundant unlabeled medical images. The local-global objectives enable strong performance on both global classification and dense structured prediction downstream tasks.

\section{Microsoft BioViL-T}
In many cases CXRs are not stand-alone. When an in-patient is undergoing treatment or being monitored for progression, radiology reports are taken daily \cite{CXRdaily}. On the other hand, patients with chronic conditions may have repeat scans every few months depending on the GP's suggestion \cite{drsharma2023interview}. For these cases, where radiologists have multiple scans at hand, they will incorporate temporal semantics in reports commenting on progression and change in any observable features. However, most existing self-supervised VLMs only consider single CXR images and their associated reports. Microsoft BioViL-T\cite{BioVil-T} explicitly account for temporal information in radiology reports by using prior images, where available, in vision language processing.

\subsection{Architecture}
BioViL-T's novel contributions includes a hybrid CNN-Transformer image encoder. The image encoder can handle up to two input images, usually DICOM images to avoid JPG suppression errors. However, the workflow of the encoder and its output embeddings vary depending on whether one or two input images are passed in. Let's first consider the case of two input images passed, with the following process:
\begin{enumerate}
    \item Both images are resized to 512 pixels and centre cropped to (448,448). Random affine transformations are applied including rotations (up to 30 degrees) and shear (up to 15 degrees) as well as colour jitter (brightness and contrast).
    \item Both images are passed through a ResNet-50 CNN which produces patch tokens of 14,14 pixels and 256 channels (i.e. each token is 14*14*256 and we have 32*32 tokens per image, overall our tensor's shape is (256,14*32,14*32).
    \item Each output token from both images is augmented with 2D sinusoidal positional encodings and learnt temporal encoding.
    \item These augmented output tokens for each image are then flattened across their spatial dimensions (i.e. creating a tensor with shape (14*32*16,14*32*16)) and both images' flattened dimensions are concatenated. 
    \item This concatenated 2D tensor is processed by K (self attention + feed-forward network)  layers, which apply cross attention over image features to capture the interactions between patch embeddings from the prior image and the current image, removing the need for image registration. These captured interactions between the prior image and the current image are aggregated into a fixed-length difference embedding.
    \item Finally the embedding of the current image from ResNet50 along with the difference embedding from the transformer are concatenated and average pooled down to a single vector. 
\end{enumerate}

In the case where no prior image is passed, a standard missing image representation embedding is used which has the same dimensionality as the difference embedding. This missing image representation is learnable and optimised during the training process.

BioVil-T also includes a BERT encoder, fine-tuned for CXRs as mentioned in section \ref{CXR-BERT}. The CXR reports are tokenised and prepended with a CLS or MLM token depending on the downstream training objective before being passed to the BERT encoder.
\subsection{Training}
There are two key training objectives: contrastive loss and masked language modelling.
Both the text and image encoder's embeddings are projected into a joint latent space using two-layer perceptron, which is optimised to maximise cosine similarity between matching text and image pairs.  In addition, previous research from Microsoft \cite{BioViL} has shown that MLM as an auxillary task stabilises and improves language understanding in multi-modal learning, since visual information can be used to disambiguate masked predictions hence cross-attention is used for MLM loss, specifically the following probability is optimised:
\begin{equation*}
    p_\theta (\text{masked token} | \text{unmasked tokens}, \text{ joint latent space embedding of image}  )
\end{equation*}

\subsection{Open source code review}
\label{fixing_bio_vil_t}
The BioVil-T codebase is available on Microsoft's HI-ML GitHub monorepo\cite{microsoft-hi-ml-repo} however, this repository does not support the latest version of PyTorch (2.1.x) because of deprecated functions when loading ResNet's pre-trained weights and a generally outdated requirements.txt for the HI-ML multimodal sub-repository. I have gone through and updated the codebase to support the latest version of PyTorch (2.1.2) as of 19/01/24. \\

The codebase released by Microsoft only supports phrase grounding using cosine similarity with a single text prompt and a single image's patch embedding. The code for patch and global embeddings to be generated from multiple images (i.e. a previous and a current image) exists, an interesting observation is that the Microsoft team only use ResNet18 with their multi-image encoder, not ResNet50. To support report generation, the codebase would have to be modified to add a text decoder element as currently the CXR-BERT model is encoder-only.

\subsection{Areas to improve/futher the SOTA}
I suggest a few angles to improve BioViL-T and its associated work on temporal semantics: 
\begin{itemize}
    \item \textbf{Improving the image encoder} by handling more input images (i.e. over 2),  replacing/augmenting ResNet with a VisionTransformer or optimising how single vs multiple images are handled

    \item \textbf{Fine-tuning}: use longitudinal, non-intensive care studies to fine-tune as BioVil-T is trained over intensive-care only MIMIC-CXR studies. Another approach could be to augment reports with GPT3.5, similar to MAIRA-1\cite{hyland2023maira-}, which could improve clinical metrics in report generation downstream tasks.
\end{itemize}

\section{BLIP-2}
\label{BLIP-2}
BLIP-2 tackles the issues of high computation cost for pre-training SOTA VLMs due to end-to-end training using large-scale datasets and models, for example, the Med PalM-M\cite{Med-PalM-M} models have between 12 - 562 billion parameters. The Salesforce team release a generic and compute-efficient vision language processing (VLP) method by bootstrapping from readily available (on HuggingFace), pre-trained vision encoders and LLMs. To connect a vision encoder's embedding output with an LLM, which has not seen images during its unimodal training, vision language alignment is required using a Querying Transformer (Q-Former).

\subsection{Q-Former}
The Q-Former is the only trainable module required, it connects a frozen image encoder and frozen LLM. It is composed of a vision transformer that interacts with the frozen image encoder for visual feature extraction and a text transformer that serves as both a text encoder and a text decoder, note however, that both transformers share the same self-attention layers. The model learns query embeddings to input to the image transformer. These queries interact with each other, the frozen image features and the text as the self-attention layers are shared. Altogether, a Q-Former contains 188M parameters including the learnt weight matrices for key, value and query vectors. There are 32 learnable query vectors in the given BLIP paper each with dimension of 768, which is far smaller than the frozen image features forcing the queries to extract visual information most relevant to the text.

\subsection{Training}

The model is trained in two stages - representation learning and generative learning.

In the representation learning stage, Q-Former is connected to a frozen image encoder and trained on image-text pairs to extract visual features relevant to the paired text. Three pre-training objectives are used:

\begin{itemize}
\item Image-Text Contrastive Learning (ITC) - This contrasts positive image-text pairs against negative pairs to align the image and text representations. The highest pairwise similarity from multiple query outputs against the text representation is considered to express the image-text similarity.

\item Image-Grounded Text Generation (ITG) - The model is trained to generate image captions conditioned on the input image. This process necessitates each query to encapsulate the entire visual feature set required to construct the text. A multimodal causal self-attention mask governs query-text interaction, directing the queries to capture essential visual details, without seeing the text.

\item Image-Text Matching (ITM) - A binary classification task to predict if an image-text pair matches or not. Hard negative mining is used to create challenging negative examples, whereby the negative pairs the model finds most difficult to separate are chosen for the binary classification task.
\end{itemize}

The second stage extends the representation learning from the frozen image encoder to include a Frozen Language Model (LLM). By integrating a fully-connected (FC) layer, the output query embeddings are linearly projected to match the LLM's text embedding dimension. These modified embeddings serve as visual prompts, streamlining the visual representation to condition the LLM output.\\

Two varieties of LLMs are deployed for experimentation: decoder-based and encoder-decoder-based. For decoder LLMs, language modeling loss is used where the LLM generates text conditioned on Q-Former's visual features. For the encoder-decoder-based LLMs, prefix language modeling loss is used, which divides a text into two parts—the prefix text concatenated with the visual representation guides the LLM's encoder, while the suffix text forms the target for the decoder's generation tasks.

The image encoders and LLMs use 16-bit float conversion during training to improve efficiency.

\subsection{Areas to improve/further the SOTA}
I suggest a few approaches to improve BLIP-2, with a focus on handling temporal semantics in radiology reports and improving diversity of training data:
\begin{itemize}

\item \textbf{Leveraging medical LLMs}: Instead of non-medical LLMs like OPT and FlanT5, using a large-scale pre-trained medical LLM like Meditron could better capture domain-specific language.

\item \textbf{Temporal-Attentive Encoder with Q-Former tuned using previous report with current indication}: Using an image encoder capable of handling multiple input radiographs over temporal sequences can substantially reduce temporal hallucinations. Fine-tuning the Q-former with previous reports and current indications could substantially improve the generated reports.

\end{itemize}

\section{CheXagent: Instruction-Tuned Foundation Model for CXR Interpretation} \label{sec:SOTA-CheXagent}

CheXagent \cite{chen2024CheXagent} applies the learnings from BLIP-2 to the CXR domain by developing an end-to-end foundational model consisting of a fine-tuned vision encoder and an LLM. The development of effective foundation models (FMs) for medical imaging, particularly chest X-rays (CXR), has been significantly challenged by issues such as the limited availability of large-scale vision-language datasets and the complex nature of medical data that many standard models are not explicitly designed to handle. Addressing these limitations, the Stanford team's core contributions include

\begin{enumerate}
    \item CheXinstruct: 6 million image-question-answer triplets for instruction tuning foundational models, spanning diverse CXR tasks like classification, detection, question answering and report generation
    \item CheXagent: an 8B parameter foundational vision-language model tuned over CheXinstruct
    \item CheXbench: an evaluation benchmark allowing for comparisons of foundational models over key CXR-related tasks
\end{enumerate}

\subsection{Architecture and training}
CheXagent assimilates three major components: a large language model (LLM) fine-tuned over general medical corpuses and radiology reports, a vision encoder for handling input CXR images, and a Q-former/bridging layer that effectively integrates the vision and language modalities.

\begin{figure}[H]
    \centering
    \includegraphics[width=0.55\textwidth]{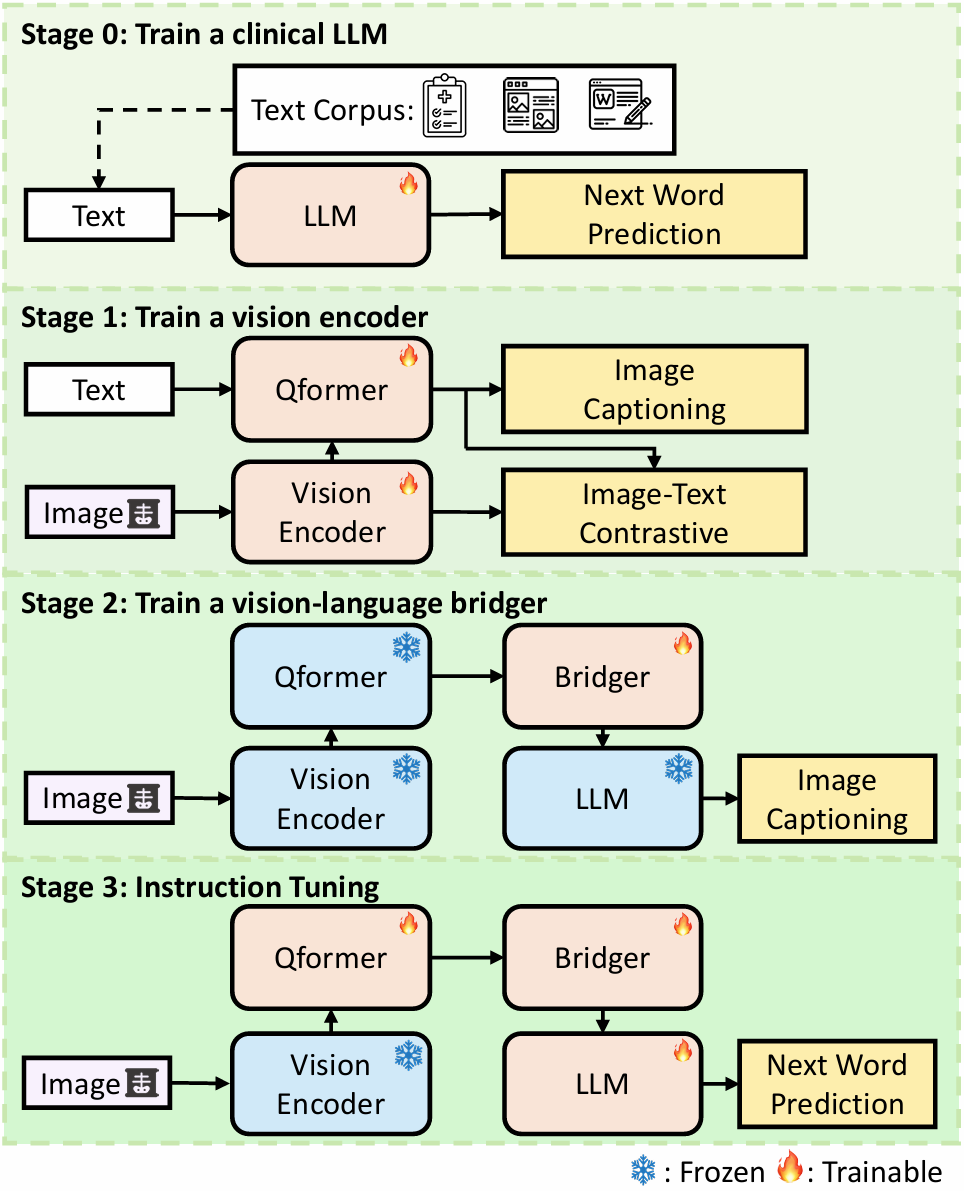}
    \caption{The four-stage training process for CheXagent}
    \label{fig:CheXagent_training}
\end{figure}

\begin{enumerate} \item \textbf{Clinical LLM Training:} The starting point is a 7B parameter LLM (Mistral-7B v0.1 \cite{jiang2023mistral}) which is adapted for the clinical domain by fine-tuning on five data sources: PubMed abstracts, MIMIC-IV radiology reports, MIMIC-IV discharge summaries, medical Wikipedia articles, and CheXinstruct CXR samples. This stage infuses the LLM with comprehensive medical and clinical knowledge, see stage 0 in figure \ref{fig:CheXagent_training}.

\item \textbf{Vision Encoder Development:}A vision encoder tailored for CXRs is trained using image-text contrastive and image captioning objectives on datasets like MIMIC-CXR, PadChest and BIMCV-COVID-19. The architecture mirrors the work of BLIP-2, using a vision transformer for visual feature extraction, see stage 1 in figure \ref{fig:CheXagent_training}.

\item \textbf{Vision-Language Bridging:} To connect the clinical LLM and CXR vision encoder, a bridging network is trained while keeping the other two components frozen. This bridges the modality gap by mapping visual data to the language space. The bridger uses an image captioning objective on the same datasets as the vision encoder, see stage 2 in figure \ref{fig:CheXagent_training}.

\item\textbf{Instructing Tuning:} Here the LLM, bridger and Q-former are fine-tuned over CheXinstruct using an instruction tuning dataset, mostly composed of MIMIC-CXR related questions, see stage 3 in figure \ref{fig:CheXagent_training}.

\end{enumerate}

\section{Med-Gemini}
We very briefly discuss Med-Gemini \cite{yang2024Med-Gemini} as it is the latest\textbf{ non-publicly available} SOTA LMM for general medical use cases, including CXR interpretation. It was released on 06/05/24, when most of this project work had been completed but our evaluation was still being conducted. We adopt aspects of their report evaluation process, in particular the report comparison rubric which clinical specialists use to compare reference radiology reports and AI-generated radiology reports. We incorporate Med-Gemini into our evaluation based on the metrics they release. \\

Gemini 1.5 \cite{gemini1.5} was fine-tuned over large-scale medical datasets to produce Med-Gemini. Specifically for chest X-rays they used MIMIC-CXR and importantly, a private dataset known as IND-1 which is of a similar scale to MIMIC-CXR but from an Indian hospital.
\chapter{An in-depth evaluation of the SOTA}
\label{chap: sota-investigation}
Having understood the architecture and training processes of state-of-the-art models used for CXR interpretation across static (single point in time scan) and longitudinal cases (many scans forming a history). The next step was to understand the performance capabilities of these models on clinically relevant tasks, with a particular focus on pathology detection, classification and localisation as recommended by Dr. Marshall, a Respiratory Specialist Registrar. \\

However, BioViL-T's released implementation did not include a decoder, as a result the model was only able to embed images and their corresponding text, hence phrase grounding was the only possible downstream task. Therefore, our investigation and evaluation primarily focus on CheXagent, the publicly available SOTA for static CXR interpretation and we only consider BioViL-T for a subset of pathology localisation tasks.

\section{A review of CheXbench for evaluating CXR FMs}
When it comes to evaluating FMs for CXR interpretation most research focuses on evaluating generated reports, however, since CheXagent is an instruction-tuned model trained over a variety of question-answer pairs it is capable of performing more tasks including view classification, disease identification, visual question answering and image-text reasoning.

Hence, the Stanford team behind CheXagent released CheXbench \cite{chen2024CheXagent} to provide a reference evaluation benchmark for a variety of tasks. CheXbench provides two evaluation axis:
\begin{enumerate}
    \item \textbf{Image perception:} 
    \begin{itemize}
        \item View classification (AP, PA, Lateral for CheXpert and an extra, fourth option LL for MIMIC-CXR )
        \item Binary Disease Classification 
        \item Single Disease Identification (single select from 4 options)
        \item Multi-Disease Identification (multiple select from 4 given options)
        \item Visual Question Answering (single select from 2 - 4 options) 
        \item Image-Text Reasoning (single select from 2 options)
    \end{itemize}

    \item \textbf{Textual Understanding}
    \begin{itemize}
        \item Findings section generation
        \item Findings summarization
    \end{itemize}

    \end{enumerate}

However, a key issue was the availability of CheXbench as it was not publicly available in late May 2024. However, we have been in contact with the Stanford AIMI team and have the Image-Text reasoning subset which we use as part of our pathology localisation evaluation framework. We do not need CheXbench to evaluate pathology detection and classification, which can be assessed using CheXpert, VinDr, and MIMIC-CXR. \\

The multiple choice question answer format used in CheXbench is not an ideal evaluation method for disease identification and vision-question answering tasks, as a radiologist is responsible not just for answering the questions outlined in an indication but also for commenting on any other abnormalities that may affect patient management. The multiple options may limit the extent to which models can comment on the subtleties since their outputs are limited to the multiple choices only. \\

Hence, we focus on linear probing and open-ended Q\&A to evaluate multi-label pathology detection and classification as outlined below in the investigation and evaluation approach.

\section{Evaluation approach}
\label{sec:evaluation_approach_sota_investigation}
Here we outline the model configurations, datasets used and metrics collected over the clinically important downstream tasks when investigating the performance of CheXagent (analysis and results from late May 2024 - subsequent iterations of the models may be released following this evaluation).

\subsection{Pathology Detection \& Classification}
\label{evaluation_approach_pathology_detection}

For pathology detection and classification, we use CheXpert-small's test set (668 scans) and a subset of VinDr's test set with 450 scans. We use a subset of VinDr's test set because the remainder is used to train our linear probes (see section \ref{Linear Probing} for an explanation). Both datasets have images with no findings, single pathologies and multiple pathologies, however each dataset has a slightly different set of pathology labels allowing for a more robust analysis of the variety of pathologies that can be detected. \\

For each scan, we compare the true labels against the predicted labels, noting there will be variation in the number of positive labels between scans with one label for a single pathology or "No finding", and multiple labels for multiple pathologies.

Our approach for evaluating the models on pathology detection and classification is as follows:
\begin{enumerate}
    \item \textbf{Open Ended Q\&A} - Prompt CheXagent with 3 different prompts:
    \begin{itemize}
        \item What pathologies are in the image?
        \item What are the findings present in the image?
        \item What abnormalities are in the image?
    \end{itemize}
    We choose these prompts as they are similar in nature to the prompts used in the CheXagent paper when evaluating foundational models on pathology detection/classification (refer to figure 7 in \cite{chen2024CheXagent}), since they span a spectrum of terminology used for this task, namely: "pathologies", "findings", and "abnormalities".
    For each of the prompts, we configure the LLM's generation configuration to have temperatures of 0.5, 1 and 1.5 to analyse whether sampling variation affects the results. \\

    \textbf{Metrics}:\\
    \label{accuracy-definitions}We calculate the accuracy, in particular:
    \begin{itemize}
        \item \textit{Exact Match Accuracy} - number of times where the pathologies predicted exactly matched the reference (i.e. no false positives or false negatives)
        \begin{itemize}
            \item We look at splits of the exact matches, where there was "No Finding", one pathology or many pathologies
        \end{itemize}
        \item \textit{Single Match Accuracy} - number of single pathologies correctly matched, including where there are multiple reference pathologies per scan and some of them may have been correctly matched, whilst others may have not been. It is possible that predicting all labels as true would lead to a good single match accuracy, in practise the models did not do this based on their exact match accuracies
        \begin{itemize}
            \item Again, we look at splits by cases where there was "No Finding" or a pathology
        \end{itemize}
    \end{itemize}
    
    \item \textbf{Linear Probing} - We analyse the performance of 4 linear probes, each probe connected to a different part of the architecture with different output neurons:
    \begin{itemize}
        \item Off of CheXagent's ViT with 22 output neurons for VinDr pathologies
        \item Off of CheXagent's Q-former with 22 output neurons for VinDr pathologies
        \item Off of CheXagent's ViT with 14 output neurons for CheXpert pathologies
        \item Off of CheXagent's Q-former with 14 output neurons for CheXpert pathologies
    \end{itemize}

    \textbf{Metrics}:\\
In addition to the accuracies mentioned above (see \ref{accuracy-definitions}), since we have access to probabilities, we can calculate the Top-K accuracy (the accuracy for the K labels with highest probabilities) and the ROC-AUC \cite{ROC-AUC}, this being the most insightful metric and one commonly seen in literature.

    \item \textbf{Comparison to reference pathology detection models} - we compare the performance of CheXagent linear probes against three models from TorchXrayVision \cite{Cohen2022xrv}, a comprehensive library known for its robustness in pathology detection and classification. The TorchXrayVision models serve as an excellent benchmark due to their training across a variety of comprehensive and diverse datasets including:

    \begin{itemize}
        \item NIH ChestX-ray14
        \item RSNA Pneumonia Detection Challenge
        \item CheXpert
        \item MIMIC-CXR
        \item Google Open Images
        \item OpenI Chest X-ray Database
    \end{itemize}

    This makes TorchXrayVision similar to CheXagent in that both models were trained over a variety of data sources and are adept at handling domain shifts that commonly occur with medical imaging data. 

    The three TorchXrayVision models we use are:
    \begin{itemize}
        \item DenseNet121 with input image size 224x224
        \item ResNet50 with input image size 512x512
        \item DenseNet121 with input image size 224x224 trained exclusively on CheXpert
    \end{itemize}
    \textbf{Metrics}:\\ 
    We collect the same metrics for comparison as we do during the linear probing.

\end{enumerate}

\subsection{Image text reasoning: Pathology Localisation}

To evaluate pathology localisation, we use the CheXbench image-text reasoning task. This benchmarked task is constructed using the OpenI dataset \cite{OpenICXRData} by selecting certain scans and constructing questions with two options for answers, for instance: "Which finding is in this chest X-ray?, elevated right diaphragm, elevated left diaphragm" with the model having to select a single option. Upon careful inspection of all the options, we see there are three types of tasks:

\begin{enumerate}
    \item Determine lateral location of pathology
    \item Determine vertical location of pathology (for instance, lower vs upper)
    \item Determine severity of pathology
\end{enumerate}

We focus only on the lateral location of the pathology. If we had more time, we would have liked to focus on vertical location too. We note that results in the CheXagent paper are based on the whole benchmark rather than just the lateral location split.
We evaluated CheXagent, at three different temperatures, on the above benchmark by constructing a prompt in the following format: f"\{question\} Option 1: \{option\_1\} Option 2: \{option\_2\}."  To evaluate BioViL-T, we explore two main approaches:
\begin{enumerate}
    \item Pass in both options (i.e. one stating left pathology and the other right pathology) to the phrase grounding model and see which has the higher activation, this will be the predicted option
    \item Pass in a single option having removed the lateral location term (i.e. elevated left diaphragm becomes elevated diaphragm) and predict the lateral location based on the coordinates with the highest activations
\end{enumerate}

For both of the above we carefully check if the activation is above 0, before making any prediction otherwise the model will just predict randomly and this will affect our results.

\subsection{Report generation}
\label{sec:report_generation_plan}

The difficulty in evaluating reports lies in the huge variation in CXR reports as different radiologists have different styles, in particular concerning brevity. Hence traditional NLP metrics to compare text such as ROUGE\cite{lin-2004-rouge} or BLEU \cite{papineni-etal-2002-bleu} only favour reports which are written in ways similar to reference reports, optimising for such metrics can result in overfitting to a particular style or convention of reporting, with domain shift present in different countries and medical settings. We do collect ROUGE-L metrics for our evaluation of the various CXR agents in section\ref{sec: report-generation-overall-evaluation}, however, this is for comparison with the clinical metrics to highlight the limitations of NLP-based metrics. \\ 

\begin{figure}[H]
    \centering
    \includegraphics[width=0.9\textwidth]{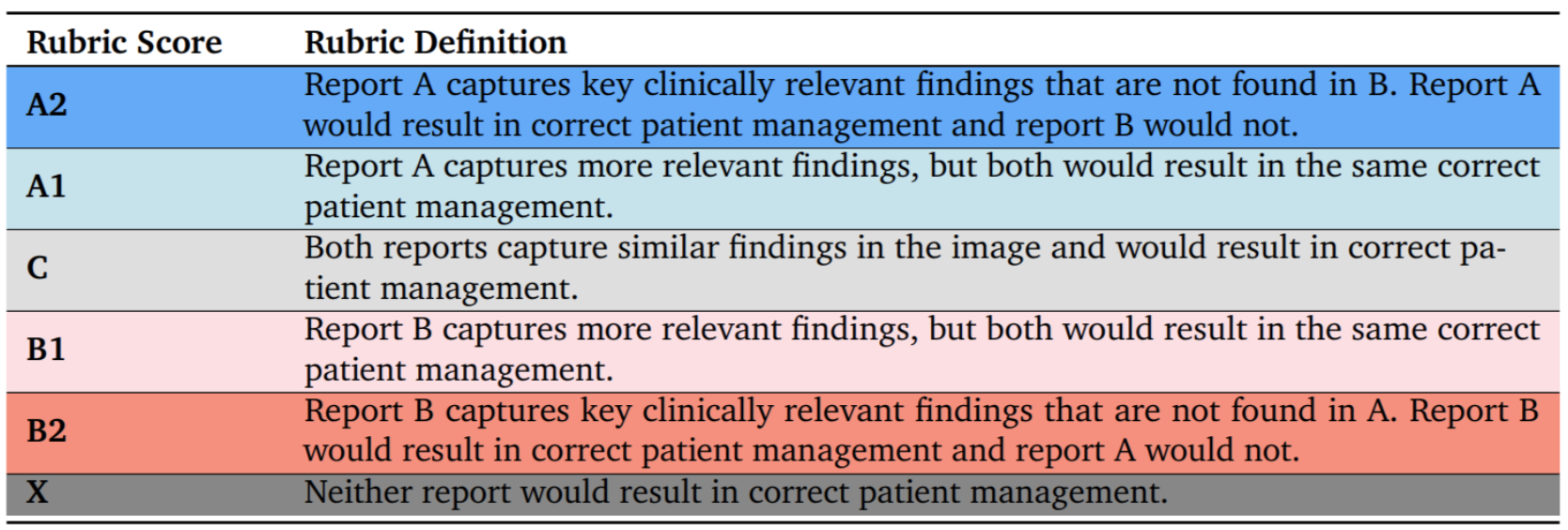}
    \caption{Google's rubric to compare two radiology reports to be used by certified radiologists}
    \label{fig:google_report_comparison_rubric}
\end{figure}

Hence, I focus primarily on clinical evaluations over MIMIC-II \cite{MIMIC-II}'s validation and test sets from which we carefully select cases with a single scan and no priors. We also evaluated performance over the CheXpert test dataset \cite{irvin2019chexpert} but without comparing it to reference reports, as these are not available. Together with certified radiologists or respiratory-specialised registrars, we evaluate CheXagent on the following 5 points:

\begin{enumerate}
    \item \textbf{Comparison to reference report:} We adopt a rubric (see figure \ref{fig:google_report_comparison_rubric}) developed by Google \cite{yang2024Med-Gemini} to compare two reports, in our case the generated reports to reference reports 
    \item \textbf{Brevity}: The clinical partner will select from \{Too Concise = -1, Good = 0, Too Verbose = +1\}, which we map to numbers allowing us to take averages and see how the reports tend to pan out in terms of verbosity.
    \item \textbf{Accuracy (1-5)}: To allow for evaluation of reports when there is no reference report, for example in the case of CheXpert, we develop an accuracy metric to assess the quality of a report based on the pathologies identified, detail of findings and impacts on patient management. Refer to table \ref{tab:accuracy_score_definition} for the rubric we used to determine accuracy scores.
    
    \item \textbf{Is report acutely dangerous?}: We add a simple option to collect data on reports that are acutely dangerous meaning that they would lead to a sudden onset with significant risk to health, for example not reporting a pneumothorax. Where accuracy score equal to 1 will account for this, we wanted to explictly flag the most dangerous reports as these indicate major failures, which would prevent adoption of the model.
    
    \item \textbf {Temporal Hallucination}: Finally if a report has any language referring to non-existent priors or general temporal language (i.e. worsening of edema), we flag this as all of our experiments are over static cases only.
    
\end{enumerate}

\begin{longtable}{>{\columncolor{gray!10}}>{\centering\arraybackslash}m{2cm} >{\centering\arraybackslash}m{12cm}}
\rowcolor{gray!20}
\textbf{Score} & \textbf{Definition} \\ \hline
5 & Perfect report, accurate detailed (no hallucinations) \\ \hline
4 & Generally accurate, a few missing details \\ \hline
3 & Key details present but requires additional interpretation with no issues regarding patient management (includes possible hallucinations) \\ \hline
2 & Missing key details - not dangerous (includes possible hallucinations) \\ \hline
1 & Dangerous (would lead to mismanagement) \\ \hline
\caption{Accuracy scoring rubric devised with Dom Marshall, respiratory-speciality registrar}
\label{tab:accuracy_score_definition}
\end{longtable}

\section{Linear probing}
\label{Linear Probing}
While the open-ended Q\&A approach allows for evaluating the model's performance in a more natural and unconstrained setting, it can be challenging to precisely quantify the capabilities of the model in specific tasks such as pathology detection, classification, and localization. For example, we are unaware of how CheXagent outputs findings it is uncertain about, meaning \textit{we cannot evaluate performance at different confidence thresholds using ROC-AUC \cite{ROC-AUC} or analyse Top-K performance}. \\ 

Hence, we use linear probing to provide a more controlled and targeted evaluation framework for assessing the model's performance and for fair comparison against baselines on these specific tasks. Linear probing also facilitates a more in-depth analysis of the model's behaviour and representations. By probing different layers or components of the model's architecture, we can gain insights into the flow of information and the representations learned at different stages of the model. \textbf{This analysis can inform future model design, refinement, and potential areas for improvement}.

\subsection{Probing pipeline}
We outline the steps required to easily and flexibly create probes off of CheXagent: 
\begin{enumerate}
    \item \textbf{Embedding extraction via hooks:} \\
    We start by adding hooks on the forward pass of CheXagent's vision encoder to collect intermediate representations generated by the Vision Transformer (embedding size of 1408) and the Q-former (embedding size of 128 x 768). 
    \item \textbf{Linear probe architecture:} \\
    The chosen linear probe was a single fully connected layer, taking in the flattened embeddings and producing output neurons mapping to the labels in our VinDr and CheXpert training sets respectively. We used a sigmoid activation to map each output neuron's activation to a range where it can be interpreted as a probability or confidence.
    \item \textbf{Probe training and hyperparameter search:} \\
    We trained the CheXpert probe over the CheXpert training set which had 20,000 images but we found that 5,000 were sufficient as with 10,000 scans our test accuracy was similar. For VinDr, the training set had 3 sets of labels per scan from different radiologists, meaning there was no consensus. Hence we trained over the test split, for which we had a single set of labels agreed upon by 5 radiologists (see VinDr paper \cite{VinDrNguyen2022} for more information on their data collection). In particular,we trained over 75\% of the test set with 2250 scans and used 10\% for the validation split and 15\% for the test split. \\
    We first prepared the training data by constructing data frames containing the unique image identifiers, stored embeddings per image collected by the hooks for both the ViT and the Q-formers and finally the ground truth pathology results for the dataset (i.e. for each pathology we would have a column with a binary label indicating its presence). We had two data frames, one for CheXpert and the other for VinDr.\\

    Our linear probe was trained using a Binary Cross Entropy loss (in practise we used BCEwithLogits and did not have a sigmoid in the training forward pass for numerical stability), our hyperparameter search spanned the following:
    \begin{itemize}
        \item \textbf{Batch Size}: [64, 128, 256, 512, 1024]
        \item \textbf{Epochs}: [10,20,40]
        \item \textbf{Learning Rate}: [0.00001, 0.0001, 0.001]
    \end{itemize}

    Evaluating each of the probes on their respective held-out test sets, we found little variance in exact match accuracy between the better hyperparameter configurations, but more than 10\% variance between the best and worst configurations. We outline the best hyperparameter configurations in the following.
    \begin{itemize}
        \item VinDr:
        \begin{itemize}
            \item Vision Transformer Probe: Batch Size = 256,  Epochs = 20, Learning Rate: 0.001
            \item Q Former Probe: Batch Size = 256,  Epochs = 20, Learning Rate: 0.0001
        \end{itemize}
        \item CheXpert:
        \begin{itemize}
            \item Vision Transformer Probe: Batch Size = 256,  Epochs = 10, Learning Rate: 0.001
            \item Q Former Probe: Batch Size = 512,  Epochs = 10, Learning Rate: 0.00001
        \end{itemize}
    \end{itemize}
    I hypothesize the far higher dimensionality and, in turn, parameter count of the Q-Former probe meant a lower learning rate was important to prevent overfitting.
    
\end{enumerate}

\section{Hypotheses}
Here, we outline our hypotheses for how we expect CheXagent to perform on each downstream task, calling out potential strengths and weaknesses where applicable.\\

\textbf{For pathology detection and classification:} we expect CheXagent's probes to perform well given the volume of labelled training data used during training of the vision encoder during training stage 1. Since the vision encoder is frozen for all later training stages we expect this learning to be retained. We suspect the probes will perform better than the full model at pathology detection and classification as they will have been tuned explictly for this task, however we are unsure how much of a difference there will be. \\

\begin{figure}[H]
    \centering
    \includegraphics[width=0.9\textwidth]{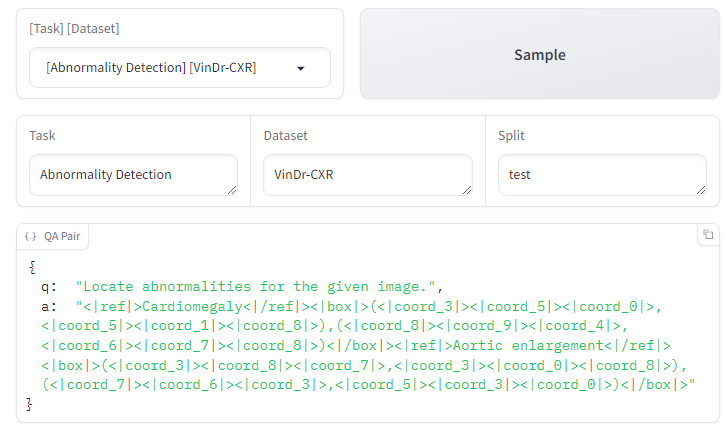}
    \caption{An example training pair for the fine-grained image understanding training task}
    \label{fig:vindr_bounding_box_training_pair}
\end{figure}

\textbf{For pathology localisation:} we do not expect great performance for a few reasons:
\begin{itemize}
    \item Training data with the lateral location of the pathologies labelled is scarce, forming around 30,000 training instances per dataset relative to other tasks which have over 1,000,000 training instances, for example MIMIC-CXR-Struct (see table 6 in appendix of CheXagent \cite{chen2024CheXagent})
    \item The image-question-answer training pairs rarely have natural language-based answers to pathologies. Instead, the answers are in the form of bounding box coordinates (see figure \ref{fig:vindr_bounding_box_training_pair} taken from the Huggingface CheXinstruct demo for example) or segmentation maps, which are not responses typically expected by radiologists on downstream tasks. Radgraph is the only clear example of natural language-based localisation of pathologies, however it has only 541 training examples.
\end{itemize}

\textbf{For report generation:} we expect to see mixed performance as the MIMIC-CXR reports used for training come from only 1 medical setting, an intensive care unit, which is not representative of the variety of CXR reports. In intensive care, most scans are longitudinal with many scans taken per patient over a short time period. Hence the language used by radiologists is often progression-related or comparative. Based on our first few inference runs of CheXagent over single scans we have already seen evidence of this overfitting with phrases such as "compared to \_\_\_ study" often appearing when we are only passing a single image with no prior scan information.\\ 

Another issue we note is the lack of uncertainty in generated reports. Generations do not caveat any observations with language to reflect confidence and mention the presence of pathologies with certainty. These hallucinated pathologies can be significantly misleading and waste the time of end users who have to manually double check for these observations.
We do not think this issue will be mitigated as it is a product of the training data, however we think about hyperparameters that could affect the generations. For example, higher temperatures could yield better performance in catching more pathologies (i.e. more false positives) due to the lower certainty allowed when sampling the next token, whereas lower temperatures would be better at catching higher certainty observations(i.e. fewer false positives, but perhaps more false negatives).

\section{Results}

We present the findings of our evaluation experiments of CheXagent, its probes and other SOTA models across the three key tasks of pathology detection and classification. In this section only, for the sake of consistency, \textit{any of the following graphs coloured blue will be based on the CheXpert data and those coloured orange will be based on the VinDr data. All data presented is the mean of 3 repeats, with variances presented where applicable.}

\subsection{Pathology Detection \& Classification}
\label{sec:pathology_detection_results}
With LMMs many prompts can be used to achieve certain tasks, therefore, when analyzing CheXagent on pathology detection and classification it was important to determine how to prompt the model. We chose three prompts as listed in the evaluation approach \ref{evaluation_approach_pathology_detection}. However, upon running CheXagent with these three prompts we noted the prompt "What are the findings in the image?" led to verbose generations which were similar to those of the findings section of a radiology report, whereas the other two prompts resulted in CheXagent simply listing the pathologies. Based on this observation and the fact we had a dedicated evaluation pipeline for evaluating findings sections of a radiology report (see section \ref{sec:report_generation_plan}), we only show results for the prompts "What pathologies are in the image?" and "What abnormalities are in the image?". \\

\begin{figure}
    \centering
    \includegraphics[width=1\linewidth]{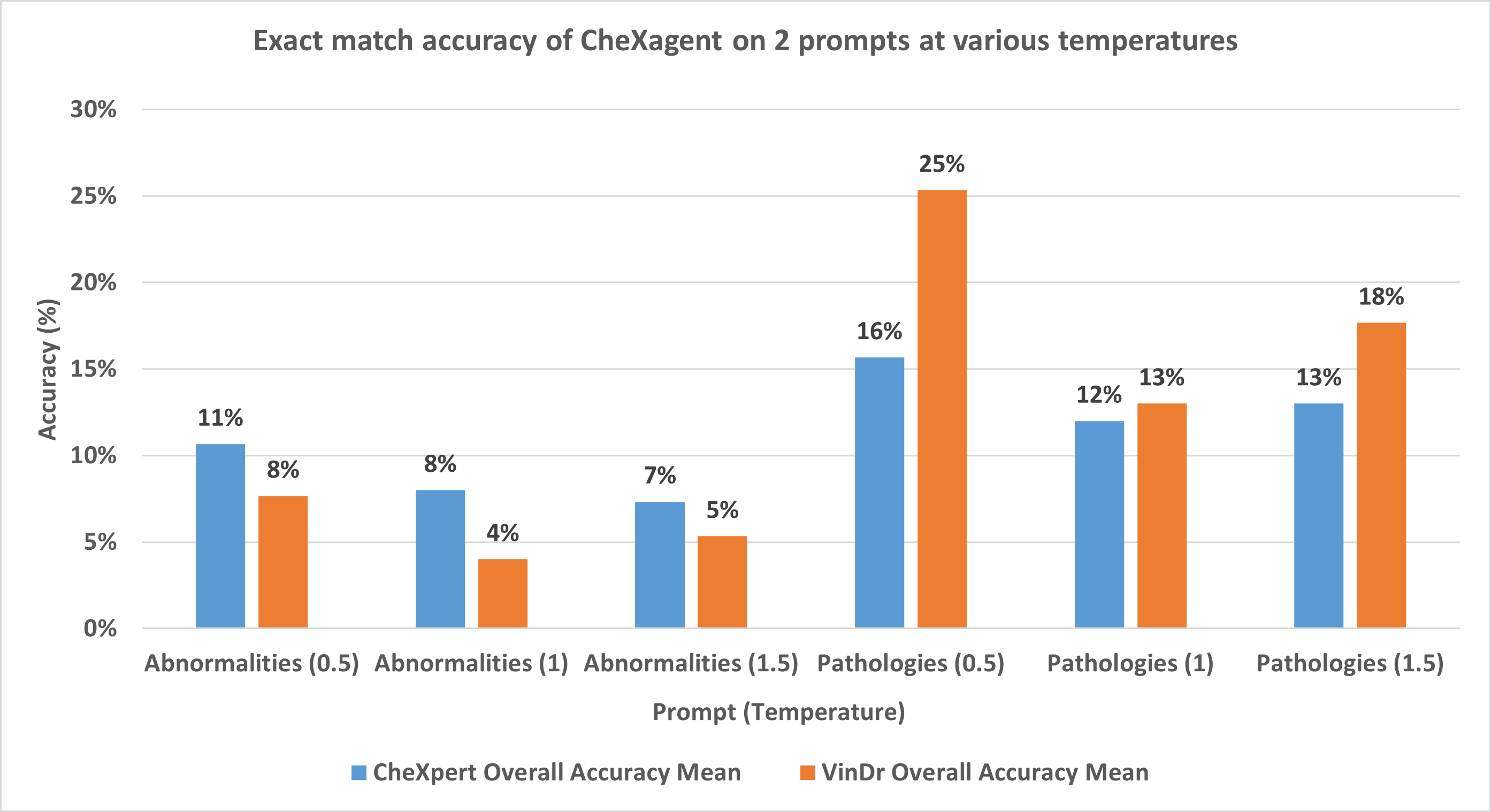}
    \caption{Comparison of pathology classification results across different prompts and VinDr/CheXpert dataset}  
    \label{fig:exact_match_CheXagent}
\end{figure}

From figure \ref{fig:exact_match_CheXagent}, we see clear variation in accuracy between the prompts with "What pathologies are in the image?" consistently scoring higher across both VinDr and CheXpert. We also note the lowest temperature sampled (0.5) results in higher accuracy for both models across both datasets. 
We find that the principal reason for the higher accuracy at a lower temperature is better performance on the "No finding" cases, with little gains in the single or multiple pathology cases.\\

Once we have determined the best prompt and temperature to use, we compare the performance of CheXagent with its probes and the Torch X-ray vision models across both datasets. We note it is likely that our prompting configuration is only a local maximum and finding the best prompt would require a far greater search space, however, this prompt is representative of what an end-user may type in when using CheXagent hence we continue with it.\\

\begin{figure}
    \centering
    \includegraphics[width=1\linewidth]{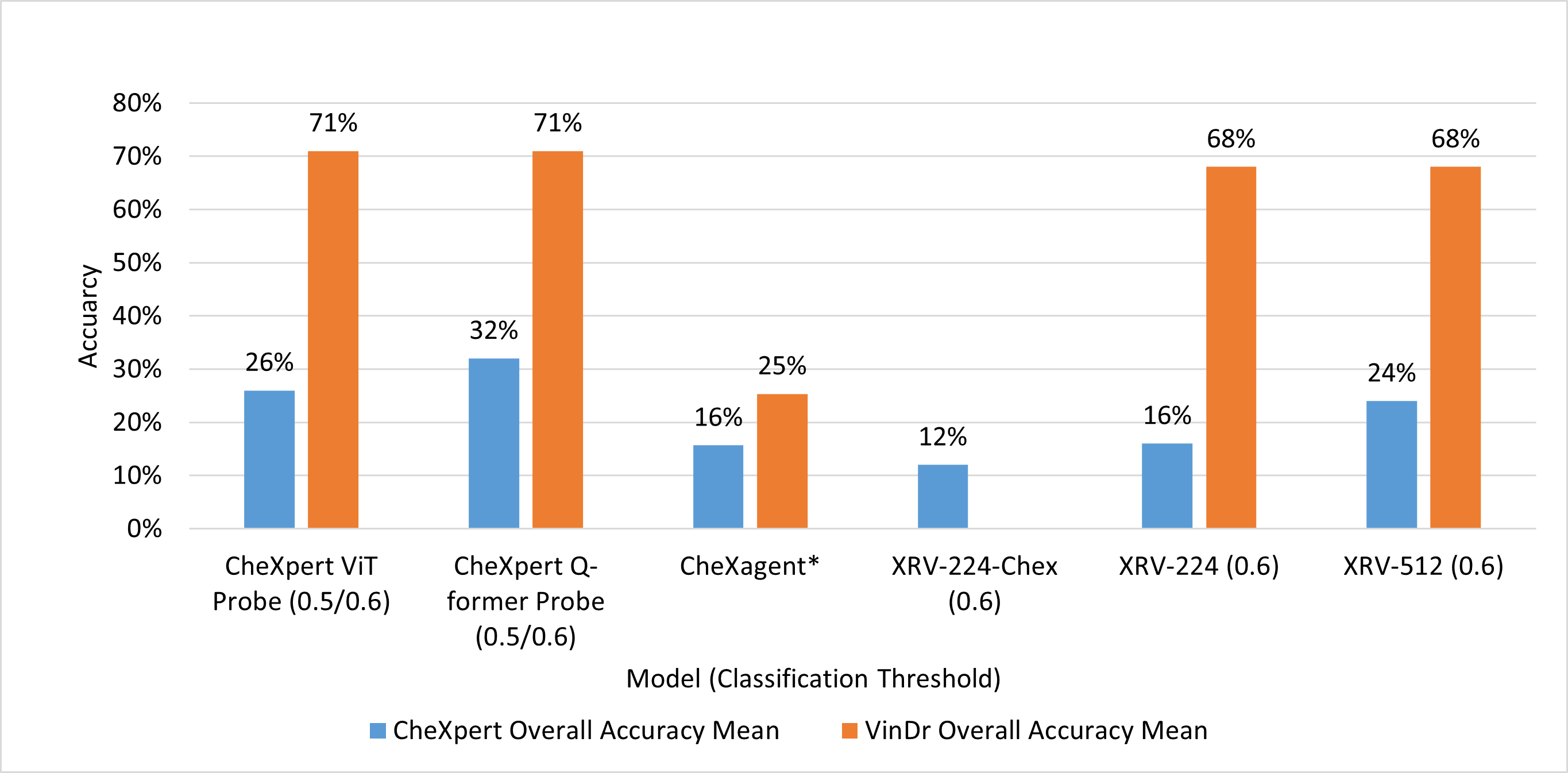}
    \caption{Comparison of pathology classification between CheXagent, CheXagent probes and Torch X-ray vision models (*: best CheXagent configuration from figure \ref{fig:exact_match_CheXagent} (pathologies prompt at temperature 0.5)) }  
    \label{fig:exact_match_CheXagent_vs_probes_vs_xrv}
\end{figure}

Based on figure \ref{fig:exact_match_CheXagent_vs_probes_vs_xrv}, for all the different models the performance on VinDr is far better than CheXpert, in many cases over twice the performance. This is possibly due to the composition of the respective datasets, with VinDr having nearly 70\% cases with "No findings" as compared to CheXpert's 25\%, suggesting these models are far better at spotting the binary case of a normal versus an abnormal scan however, they struggle more with the complex abnormal cases where there are many pathologies, noting CheXpert has 57\% of its dataset consisting of these cases.\\

Figure \ref{fig:exact_match_CheXagent_vs_probes_vs_xrv} shows that performance of the probes is far higher than the overall CheXagent model and the other publicly available Torch X-ray Vision models. The strong performance of the probes indicates that the latent understanding of the CheXagent model is very robust. However, this does not reflect in the overall output from its LLM, suggesting some bottlenecking or insufficient attention over the vision transformer/Q-former embedding spaces. We hypothesize that this may be due to the limited amount and diversity of report-based training data available to fine-tune CheXagent's LLM. This lack of sufficient text-based training data might be preventing the model from fully leveraging the rich embeddings produced by the vision transformer/Q-former. Addressing this bottleneck could involve augmenting the training data with more diverse and detailed medical reports, as well as refining the integration mechanisms between the visual and language components of the model.\\

Often when comparing models which return probabilities per pathology it is hard to determine a threshold to compare the performance, hence ROC-AUC provides a better insight into the latent performance of these classifiers across various thresholds and Top-K allows to compare performance by looking at each classifier's perception of relative confidence across the pathologies irrespective of absolute confidence.This is crucial in clinical settings where the model's top suggestions are more likely to be reviewed by healthcare professionals. By examining the Top-K accuracy, we can determine how often the Top-K predictions are included in the reference pathologies, providing a practical measure of the model's utility in real-world diagnostic scenarios. We only evaluate over K=1 since many cases have either "No finding" or a single pathology hence for K \textgreater 1, we would be overpredicting.\\

From table \ref{tab:model_performance} we can see that the ViT and Q-former probe results are very similar across both datasets in terms of ROC-AUC and Top K=1. The ViT probe, in particular, is better than all other XRV models across both datasets on both metrics showing its SOTA performance relative to other classifiers. This table once again highlights the relative complexity of the datasets again with all models achieving far lower scores on CheXpert than VinDr, with the exception on XRV-224's Top K result. Whilst the performance does vary between datasets, the difference is by far greatest for XRV-512, which does not perform well on the difficult CheXpert dataset, suggesting some shortcomings in tackling domain shifts perhaps because the model has been undertrained, especially when one considers the higher parameter count due to the larger encoder input size.

\begin{longtable}{>{\columncolor{gray!10}}>{\centering\arraybackslash}m{4cm} >{\centering\arraybackslash}m{2.5cm} >{\centering\arraybackslash}m{2.5cm} >{\centering\arraybackslash}m{2.5cm} >{\centering\arraybackslash}m{2.5cm}}

\rowcolor{gray!20} 
\textbf{Model} & \textbf{CheXpert ROC-AUC} & \textbf{VinDr ROC-AUC} & \textbf{CheXpert Top K=1} & \textbf{VinDr Top K=1} \\ \hline
\endfirsthead

\rowcolor{gray!20} 
\textbf{Model} & \textbf{CheXpert ROC-AUC} & \textbf{VinDr ROC-AUC} & \textbf{CheXpert Top K=1} & \textbf{VinDr Top K=1} \\ \hline
\endhead

ViT Probe &\textbf{ 0.853} &\ 0.958 & \textbf{0.66 }& 0.83 \\ 
Q-Former Probe & 0.832 & \textbf{0.963} & 0.64 & \textbf{0.86} \\ 
XRV-224-Chex & 0.815 & n/a & 0.49 & n/a \\ 
XRV-224 & 0.841 & 0.870 & 0.51 & 0.45 \\ 
XRV-512 & 0.620 & 0.912 & 0.27 & 0.68 \\ 

\caption{Performance comparison of classifiers on CheXpert and VinDr datasets.}
\label{tab:model_performance}
\end{longtable}

\subsection{Image Text reasoning: Pathology Localisation}
\label{sec:image text reasoning results}

From table \ref{tab:localisation_accuracy}, we see a variation by 10\% in localisation accuracy for CheXagent highlighting the sensitivity of the model to its generation configuration, interestingly the worst performance comes at the lowest temperature unlike with pathology detection and classification.\\

BioViL-T demonstrates the best pathology localisation accuracy, however this is relative to the options for which it recorded a non-zero activation, 248 non-zero activations out of 293 left vs right test examples. On this subset, we compare CheXagent's best configuration at temperature 1 and find it achieves only 57\%. This means that whilst CheXagent may be able to localise on a wider range of medical observation than BioViL-T, its accuracy is lower across the same common set of observations. Although based on the 11\% drop in BioViL-T's performance without the left/right option in the phrase grounding prompt, we realise this location descriptor is imperative for good performance. We believe the more detailed the phrase being embedded by BioViL-T's fine-tuned BERT model, the richer the embedding representation and the better the phrase grounding through cosine similarity.

\begin{longtable}{>{\columncolor{gray!10}}>{\centering\arraybackslash}m{10cm} >{\centering\arraybackslash}m{5cm}}
\rowcolor{gray!20}
\textbf{Model} & \textbf{Localisation Accuracy} \\ \hline
\endfirsthead
\rowcolor{gray!20}
\textbf{Model} & \textbf{Localisation Accuracy} \\ \hline

\endhead

CheXagent (Temp=0.5) & 45\% \\
CheXagent (Temp=1) & 55\% \\
CheXagent (Temp=1.5) & 53\% \\\\ \hline
BioViL-T (CheXbench options unchanged over non-zero activations) & \textbf{62\%} (CheXagent* same subset: 57\%) \\\\ 
BioViL-T (CheXbench options unchanged over all) & 52\% \\ \hline
BioViL-T (removing only left/right from each CheXbench option over non-zero activations) & 51\% \\\\
BioViL-T (removing only left/right from each CheXbench option over all) & 37\% \\ \hline

\caption{Localisation Accuracy of different models on CheXbench's image text reasoning split. CheXagent* = run at temperature 1}
\label{tab:localisation_accuracy}

\end{longtable}

\subsection{Report Generation}
\label{sec:CheXagent_report_generation}
Firstly, thank you to Dr Dominic Marshall and Dr Yueqi Ge, the respiratory registrars who helped with these evaluations. Here we present the results of the clinical evaluation of CheXagent's generated reports.
A key point to note when interpreting table \ref{tab:CheXagent_report_eval_metrics} is that we map the Google rubric scores to numbers for easy comparison, specifically we apply the following mapping \{X: 0, B2: -2, B1: -1, C: 0, A1: 1, C2: 2\}, given that we consider A to be when the AI generated report is better, a higher score for this metric is desired

\begin{longtable}{>{\columncolor{gray!10}}>{\centering\arraybackslash}m{6cm} >{\centering\arraybackslash}m{2.5cm} >{\centering\arraybackslash}m{2.6cm} >{\centering\arraybackslash}m{2.5cm} >{\centering\arraybackslash}m{2.5cm}}
\rowcolor{gray!20}
\textbf{Evaluation metric} & \textbf{By Normal scan} & \textbf{By Abnormal scan} & \textbf{Overall} \\ \hline

\endfirsthead

\rowcolor{gray!20}
\textbf{Evaluation metric} & \textbf{By Normal scan} & \textbf{By Abnormal scan} & \textbf{Overall} \\ \hline

\endhead

Reference report comparison $\uparrow$ & -0.1 & -1.24 & -0.91 \\
Brevity* & -0.05 & 0 & -0.01 \\
Accuracy $\uparrow$ & 4.5 & 2.44 & 3.03 \\
Dangerous report count $\downarrow$ & 0 & 4.5 & 4.00 \\
Reports with temporal hallucinations $\downarrow$ & 0 & 11 & 11.00 \\

\caption{Evaluation metrics for the report generation task over 10 normal and 25 abnormal \textbf{MIMIC-II} scans interpreted by Dom and Yueqi separately with an average taken. * for brevity, 0 is best with -1 being too concise and 1 being too verbose}
\label{tab:CheXagent_report_eval_metrics}

\end{longtable}

\begin{longtable}{>{\columncolor{gray!10}}>{\centering\arraybackslash}m{6cm} >{\centering\arraybackslash}m{2.5cm} >{\centering\arraybackslash}m{2.6cm} >{\centering\arraybackslash}m{2.5cm} >{\centering\arraybackslash}m{2.5cm}}
\rowcolor{gray!20}
\textbf{Evaluation metric} & \textbf{By Normal scan} & \textbf{By Abnormal scan} & \textbf{Overall} \\ \hline

\endfirsthead

\rowcolor{gray!20}
\textbf{Evaluation metric} & \textbf{By Normal scan} & \textbf{By Abnormal scan} & \textbf{Overall} \\ \hline

\endhead

Brevity* & 0 & 0 & 0.00 \\
Accuracy $\uparrow$ & 4.82 & 1.84 & 2.93 \\
Dangerous report count $\downarrow$ & 0 & 6 & 6.00 \\
Reports with temporal hallucinations $\downarrow$ & 0 & 15 & 15.00 \\

\caption{Evaluation metrics for the report generation task over 11 normal and 19 abnormal \textbf{CheXpert} scans interpreted by Dom. * for brevity, 0 is best with -1 being too concise and 1 being too verbose }
\label{tab:CheXagent_report_eval_metrics_chexpert}

\end{longtable}

From tables \ref{tab:CheXagent_report_eval_metrics} and \ref{tab:CheXagent_report_eval_metrics_chexpert}, we see a clear disparity in CheXagent's performance on normal scans versus abnormal scans. Across every single metric apart from brevity, CheXagent's performance on normal scans is better than abnormal scans. For normal scans, respiratory doctors prefer concise reports which call out certain anatomies (i.e. heart size, clear lungs, ...). When looking at CheXagent's generated reports for normal scans these are all very similar (almost the exact same); however, they are in line with what respiratory doctors expect. \\

However on abnormal scans, CheXagent's performance is poor with many temporal hallucinations (nearly half of all abnormal reports), missed pathologies, and in some cases, acutely dangerous reports. Based on the commonly occurring phrase "In comparison with the study of \_\_\_" and references to nonexistent pathologies/support devices, we believe overfitting and a lack of variation in reports to be a key factor for this poor performance. This is understandable given that the only reports used to train CheXagent came from a single hospital's ICU which is a very limited and not highly representative medical setting. \\

From sitting with experts during the evaluation process spanning 6+ hours, we learnt that an implication of concise reports or reports with hallucinated findings was the added interpretation time to check if the registrars agreed with the generated finding. In particular, CheXagent often confidently hallucinated in scans which resulted in slower interpretations and lower trust in the technology. Even though the model's outputs were conducted blind (refer to section \ref{sec:evaluation_data_collection_platform}, evaluators quickly noticed the style of CheXagent's reports and had preconceptions regarding its performance.

\section{Key Takeaways}
\label{sec:key_takeaways}
We conclude our investigation of CheXagent and BioViL-T by presenting our key takeaways, which motivates our implementations and contributions presented in Chapter 5.

\begin{itemize}
    \item In general, all models tested are better at distinguishing between normal and abnormal scans as compared to correctly identifying all pathologies in complex multi-pathology scans, common in an ICU setting. Refer to section \ref{sec:pathology_detection_results} and figure \ref{fig:exact_match_CheXagent_vs_probes_vs_xrv}, and section \ref{sec:CheXagent_report_generation} and table \ref{tab:CheXagent_report_eval_metrics}.
    
    \item CheXagent's performance on pathology detection and classification varies considerably based on the prompt and generation temperature. Lowest temperatures are the best primarily because of better performance on cases with no pathologies. Refer to section \ref{sec:pathology_detection_results} and figure \ref{fig:exact_match_CheXagent}.
    
    \item CheXagent's lack of confidence values for a pathology and lack of confidence-based language in radiology reports reduces model interpretability. Respiratory doctors use the radiologist's choice of language regarding pathologies to guide their interpretation, allowing for better patient management. However, the confident mispredictions in CheXagent's reports result in greater time spent interpreting the X-ray as the doctor will have to carefully double-check if they are unsure or disagree with the report. Refer to section \ref{sec:CheXagent_report_generation} and table \ref{tab:CheXagent_report_eval_metrics}.

    \item Both CheXagent probes perform considerably better than the end-to-end CheXagent model, and the Torch X-ray vision models, on pathology detection and classification across both datasets. This is likely due to the volume and diversity of data over which the ViT and Q-former are trained in this foundational model. These probes return confidence values for each pathology allowing for better interpretability. Given this result, it is likely that LLM in these large multi-modal models bottlenecks pathology detection performance. Refer to section \ref{sec:pathology_detection_results}, figure \ref{fig:exact_match_CheXagent_vs_probes_vs_xrv} and table \ref{tab:model_performance}.

    \item BioViL-T shows superior lateral pathology localisation to CheXagent on the pathologies/phrases it can ground however, for optimal performance it requires more detailed phrases, for instance the phrase to be prepended with a position (i.e. left raised diaphragm) before applying the phrase grounding's cosine similarity. We hypothesize the more detailed the phrase being embedded by BioViL-T's fine-tuned BERT encoder, the richer the embedding representation and the better the phrase grounding. The CheXagent LLM may be bottlenecking this localisation performance, however given the close-ended nature of the evaluation with the task being a binary selection task, we think this bottlenecking will be less significant than in open-ended questions, hence we believe BioViL-T to be better. Refer to section \ref{sec:image text reasoning results} and table \ref{tab:localisation_accuracy}.

    \item CheXagent hallucinates frequently, especially by referring to non-existent priors. This indicates a degree of overfitting or insufficiently diverse training data. The simplest solution would be to collect more paired scans and reports (in the order of 100,000s or millions), ideally from static cases across many medical settings. However, an alternative is to perform report augmentation (i.e. using GPT4) to increase the variety in report styles and remove references to non-existent priors. Refer to section \ref{sec:CheXagent_report_generation} and table \ref{tab:CheXagent_report_eval_metrics}.

\end{itemize}

\chapter{Agent-based CXR report generation}
\label{chap: cxr-agent}
In this chapter, we outline how we developed an agent-based CXR report generation tool based on our key observations from the evaluations in chapter 4 (refer to section \ref{sec:key_takeaways}).

\section{Objectives}
We aim to achieve the following objectives for our report generation tool:
\begin{itemize}
    \item \textbf{Uncertainty-aware radiology reporting:} Improve interpretability by propagating pathology and localisation confidence to the end-user
    \item \textbf{Improve localisation capabilities} by passing possible observations through phrase grounding tools
    \item \textbf{Minimise hallucinations} by carefully passing data to the LLM and through stringent prompt engineering
    
\end{itemize}

\section{Architecture}
To achieve the above objectives and based on the limitations of the publicly available data, we do not try to train an LMM/VLM end-to-end. This would require many radiology reports, in the order of 100,000s to 1,000,000s from a variety of medical settings. We believe this to be true based on the SOTA performance Google achieved with their Med-Gemini models demonstrating benefits of using richer, diverse training data to fine-tune Gemini \cite{yang2024Med-Gemini}. They used IND-1, a dataset of similar scale to MIMIC-CXR but from an Indian hospital, in addition to MIMIC-CXR. As of late May 2024, Med-Gemini is not publicly available hence it does not feature in our experiments or evaluations.\\

We instead focus on using the available CheXagent model and training a probe off its ViT, then passing the data from this probe to an LLM, which end users can prompt to generate the findings section of a radiology report. This allows us to benefit from the SOTA-level performance of the foundational ViT in CheXagent, whilst retaining a notion of confidence and propagating this confidence through the system to the end user.\\

To achieve pathology localisation, we use BioViL-T as a phrase grounding tool which indicates where in the image certain phrases are likely to be based on the cosine similarity of patch embeddings and the phrase embedding. In an ideal world, we would prefer to use probes off a foundational model but our early results of probing CheXagent on these tasks showed limited performance.

\begin{figure}[ht]
  \centering
  \adjustbox{lap=-0.1\linewidth}{\includegraphics[width=1.2\linewidth]{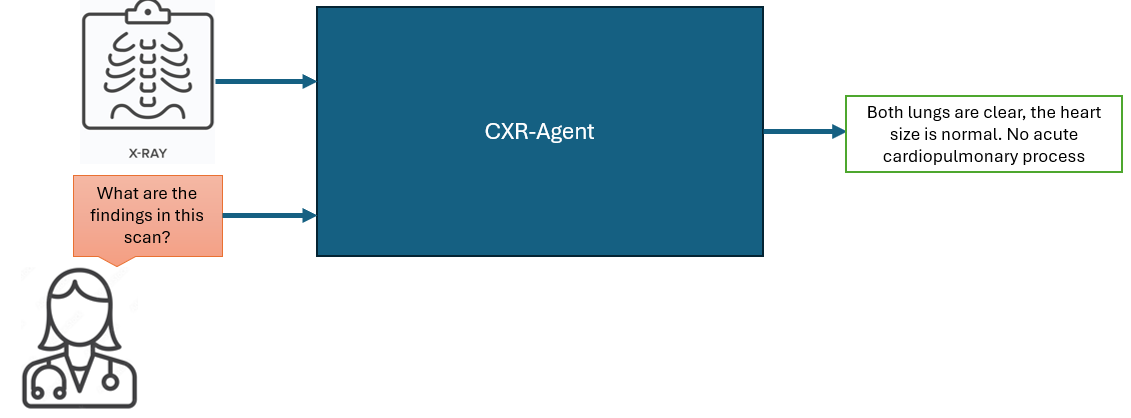}}
  \caption{A high level overview of how an end user can interact with CXR agent}
  \label{fig:cxr-agent-simple-architecture}
\end{figure}

\begin{figure}[h!]
  \centering
  \adjustbox{lap=-0.15\linewidth}{\includegraphics[width=1.3\linewidth]{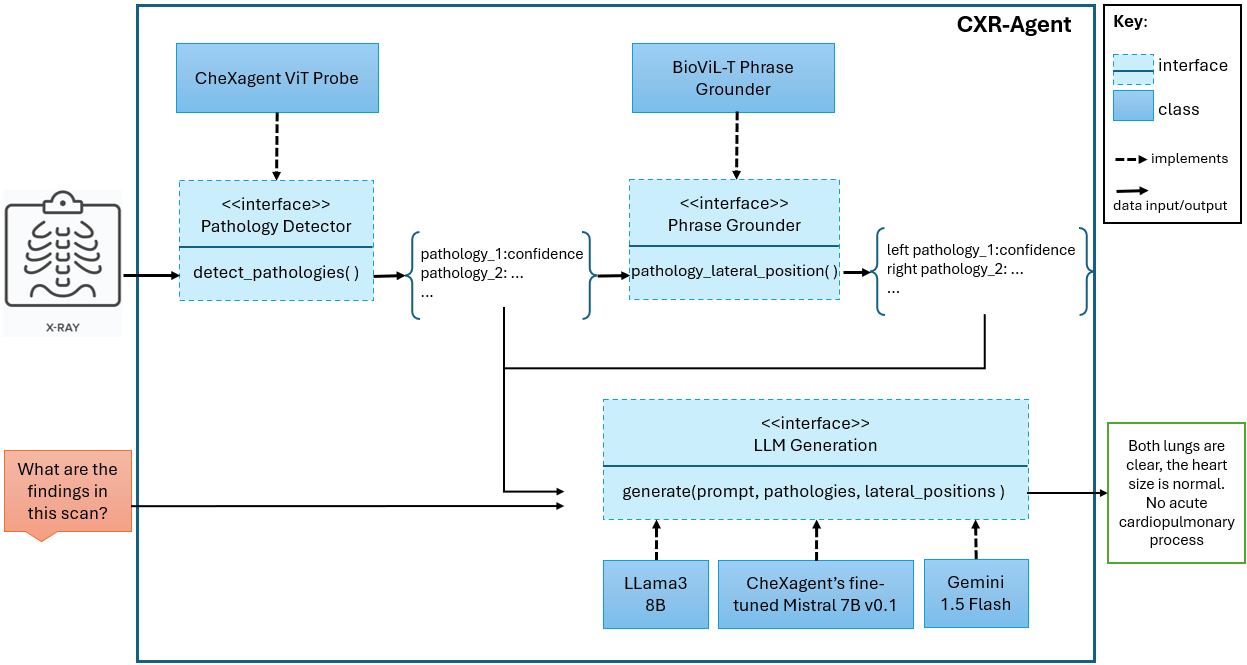}}
  \caption{A detailed system architecture of CXR-agent showing the key interfaces, concrete classes and data flow.}
  \label{fig:cxr-agent-detailed-architecture}
\end{figure}

\subsection{Architectural design choices:}
Figure \ref{fig:cxr-agent-detailed-architecture} shows the architecture of our CXR agent for which we make the following key design choices:
\begin{itemize}
    \item \textbf{Modular design}: keeping the pathology detection, phrase grounding and LLM as separate components allows us to easily change our setup in terms of removing/adding information flowing to the LLM, for instance we can easily remove the phrase grounding.
    \item \textbf{Polymorphic and flexible}: by defining clear interfaces we can easily swap out between different concrete implementations of pathology detection, phrase grounding and LLMs. We make use of this by testing between 3 different LLMs for the final findings generation.
    \item \textbf{Highly controllable}: we choose to pass the output from both the pathology detector and the phrase grounder to the LLM separately, which allows us to control the mapping of the confidence thresholds to natural language independently for each component (important since models have differences in absolute confidences). It also allows us to control which pathologies are not localised but are detected, for instance cardiomegalies. 
\end{itemize}

\section{Pathology Detection \& Classification}

For the pathology detection and classification, we chose to use the trained probe off of CheXagent's ViT, for three key reasons:
\begin{enumerate}
    \item \textbf{Performance}: The ViT probe beat all other XRV models on every single evaluated metric on both datasets and outperformed the Q-former probe on the more complex CheXpert dataset across both ROC-AUC and Top K=1.
    \item \textbf{Faster inference}: CheXagent's ViT achieves faster inference than the Q-former, mostly because the Q-former requires the ViT's outputs as inputs (meaning it will always be slower than the ViT) but also because the Q-former has the additional overhead of cross-attention mechanisms.
    \item \textbf{Compact embeddings}: Whilst both Q-former and ViT probes produce the same number of output labels depending on the pathologies they are detecting, the ViT's output is an embedding of size 1024 whereas the Q-former's output is an embedding of (128,768) meaning the linear probe has to have far more parameters, resulting in more complexity and potentially greater risk of overfitting given both probes are trained across the same two relatively small training datasets ($\leq$ 5000 training samples).
\end{enumerate}

At inference time, we pass a test scan through CheXagent and modify the forward pass to take a flag that returns with ViT embeddings output before passing through our probe's trained linear layer. We vary the probe depending on which pathologies we want to detect, however since our evaluations were on MIMIC and CheXpert we chose to use the CheXpert probe to avoid the challenges of label fusion/equivalence. We then pass the outputs of this probe to the LLM as a dictionary with pathology mapping to confidence. \\

Working with respiratory registrars, we set up thresholds to map confidence probabilities to natural language as this is far easier to interpret for our end-user. We also ignore the support devices label as this is too vague in the context of a radiology report, in which radiologists would comment on specifics for instance tube positioning. 

We applied the following mapping, noting that we do not comment on any pathology with confidence below 0.3:

\begin{table} [h]
    \centering
    \begin{tabular}{|c|c|}
        \hline
        \rowcolor{gray!20}
        \textbf{Confidence probability} & \textbf{Phrase for report} \\
        \hline
        \(0.3 \leq x < 0.5\) & "cannot exclude \textless{}pathology\textgreater{}" \\
        \(0.5 \leq x < 0.7\) & "possible \textless{}pathology\textgreater{}" \\
        \(0.7 \leq x < 0.9\) & "probable \textless{}pathology\textgreater{}" \\
        \(0.9 \leq x\) & "there is \textless{}pathology\textgreater{}" \\
        \hline
    \end{tabular}
    \caption{Working with respiratory specialists looking at data from multiple datasets we applied the above mapping from probe confidences to natural language. }
    \label{tab:probability_to_radiology_language}
\end{table}

\section{Pathology Localisation}
Based on the observations from section \ref{sec:image text reasoning results}, we see that BioViL-T outperforms CheXagent for pathology localisation, particularly when more information is supplied with the pathology for phrase grounding. We do not explore setting up probes for CheXagent localisation because of time constraints and the fact that BioViL-T's phrase grounding acts as a more general tool that could be used in wider agent-based workflows including reducing hallucinations in generated reports. To use BioViL-T, we had to fix issues with the Microsoft HI-ML repository (see section \ref{fixing_bio_vil_t}), we submitted our fixes to Microsoft, which they have since actioned.\\

For best phrase grounding performance, we maximise information being passed to the BioViL-T phrase grounding tool by prepending location descriptors, namely "left" or "right" to the pathologies detected by the pathology detector. For each pathology we get a location and a confidence, both of which we pass to the generation engine to synthesize. We acknowledge by having two different models in the system, the confidence values produced by the models will not be comparable hence we clearly mention this in the system prompt to the LLMs. Ideally, we would have a single model for the detection and localisation, which may be possible if probes for localisation work as these implicitly indicate presence.

\section{LLM Generation}
\label{sec:generation_engines}
The final component of our architecture is the LLM which synthesizes all the information from the pathology detector and the phrase grounder to generate the findings section of the radiology report. In particular, we evaluate three different LLMs to determine the characteristics that lead to higher quality of generated findings sections. We do not use the same prompt for each model rather we carefully adjust the prompts to maximise the quality of the report and ensure it hits the objectives outlined earlier on uncertainty-awareness, localisation and hallucinations. Refer to `base-agent/generation\_engine.py` in the provided codebase to see the exact prompt for each model. Note, prompts discussed below prepended to the end-user's prompt and can be thought of as image context prompts.\\

Below we list our general observations when prompting these models:
\begin{itemize}
    \item The system prompt is a very powerful way to get the model to action a specific instruction or style. Instructions that were often ignored in the image context prompt, when added to the system prompt would yield great results. (The image context prompt refers to the instruction prompt which we construct using information from the pathology detector and phrase grounder. We append the end-user's prompt to this image context prompt at inference time).
    \begin{itemize}
        \item This is the system prompt we used for the Gemini 1.5 Flash model: "You are a helpful assistant, specialising in radiology and interpreting Chest X-rays. Please answer CONCISELY and professionally as a radiologist would."
    \end{itemize}
    
    \item In the case of no pathologies being detected, we found having a custom system and image context prompt allowed for concise clear reporting.  These prompts were different to the prompts used when pathologies were detected.
\end{itemize}

\subsection{CheXagent fine-tuned Mistral 7B LLM}
We chose to include the CheXagent fine-tuned Mistral 7B model to leverage its specialized medical knowledge (refer to section \ref{sec:SOTA-CheXagent}) and compare its performance relative to CheXagent, given that that the ViT and LLM components are essentially the same but assembled differently. Our early interactions showed this model was highly sensitive to complex prompts, so we kept the prompts as simple as possible and avoided feeding thresholds into the LLM. In particular, we noted that adding a 1-shot example ("Here is a model example, 'There is a probable Pleural Effusion, and it is possibly on the left side' " ) did not help guide the style of the output rather the model would always include the example phrase hence we did not adopt any n-shot prompting for this approach.

\subsection{Llama 3 8B}
The purpose of evaluating Llama 3 8B was to see how a similar-sized, open source, SOTA LLM would perform with thorough prompting. Llama 3 was considerably better than CheXagent's fine-tuned Mistral LLM at following our instructions to produce the findings section of the report. This demonstrated that even though the model did not have any specific fine-tuning applied over the CXR domain, it had medical understanding from its pretraining alone. However, Llama 3 required a very long prompt with many specific instructions to improve its generation quality, in particular, we provided the following key pointers in the prompt:
\begin{itemize}
    \item A pathology and its lateral location (e.g., Pleural Effusion and left Pleural Effusion) are part of the same finding. The location attribute is an additional detail about where the pathology is likely found, not an indicator of a separate pathology. 
    \item Synthesize the pathology detection and localization data. Do not talk about them separately. 
    \item Confidence scores from the pathology detection and phrase grounding tools are not directly comparable. They serve as indicators of confidence within their respective contexts of pathology detection and localisation.
    \item A missing lateral location does not imply the absence of a pathology; it indicates the localisation could not be confidently determined.
\end{itemize}

Most significantly, we noted that Llama 3's generations were consistently too verbose which were only tackled by a strong intervention in the system prompt using the following phrase: "You MUST answer CONCISELY and professionally as a radiologist would." We had to add and capitalise the word "must" for Llama 3's brevity to increase. Additionally, we found Llama3 would not always comment on all pathologies detected and its localisation performance was limited as compared to Gemini 1.5-Flash as it did not pick up on the desired style and would overfit when given an example.

\subsection{Gemini 1.5-Flash}
The Gemini 1.5-Flash model was evaluated to assess how a slightly larger but still relatively small LLM performed on the task. Despite being a closed-source LLM, it is one of the most affordable options on the market, offering low latency and high throughput, making it a viable choice for real-world implementations. Gemini 1.5-Flash excelled in following the prompts and generally did not require much tailoring, producing consistent and high-quality reports. However, it also did not always use all the information on pathology location provided in the image context prompt, albeit less frequently than Llama 3.
\chapter{Evaluation of CXR-agent}
\textbf{Please note a considerable amount of evaluation is conducted in Chapter \ref{chap: sota-investigation}, which helps motivate the key objectives and design of the CXR-agent.}
\section{Evaluation Approach}

Our evaluation approach is similar to that laid out in section \ref{sec:evaluation_approach_sota_investigation}, insofar as we evaluate CXR-agent over pathology detection \& classification, pathology localisation and clinical report evaluation.\\

For pathology detection \& classification, we only evaluate on CheXpert and VinDr's test splits collecting the "Exact Match Accuracy" and "Single Match Accuracy" as defined earlier in the metrics section (\ref{evaluation_approach_pathology_detection}). We show splits of these metrics by no finding, single pathology and multiple pathology, where applicable. Unlike section \ref{sec:pathology_detection_results}, we cannot collect ROC-AUCs or Top-K metrics because our method maps confidences to natural language to improve clinical interpretability. The challenge with using natural language lies in extracting the detected pathologies. Since these models return sentences instead of a list of pathologies, we apply a regex to extract the pathologies from the output before comparing them to the reference pathologies. We first include all the pathologies in the text. Then the regex processes each sentence and checks which pathologies stated are absent based on negation keywords. We ensure to search for pathologies following these negation keywords, which may be connected by "or" or "and", such as in the phrase "no pleural effusion or opacity".
\begin{verbatim}
    sentences = re.split(r'(?<!\w\.\w.)(?<![A-Z][a-z]\.)(?<=\.|\?)\s', text)
    negation_keywords = r"\bno\b|\bnot\b|\bwithout\b|\babsent\b"
\end{verbatim}

For pathology localisation, we evaluate across CheXbench's image text reasoning split but only on the subset of tasks that require the model to determine the lateral location of the pathology. We do not focus on severity or vertical position of the pathology, though we note that a higher detection confidence usually means a more severe pathology. Testing this hypothesis with empirical results could form an interesting extension. We do not modify any of the prompts from the task, we just structure the user prompt as 'f"\{question\} Option 1:\{option\_1\} or Option 2: \{option\_2\}."' for example: "Which finding is in this chest X-ray?  Option 1: left lung opacity or Option 2: right lung opacity". \\

Our primary focus of CXR-agent's evaluation is on its report generation capabilities. We work with two experienced respiratory-specialist registrars to evaluate the generated reports from each generation engine in section \ref{sec:generation_engines} relative to CheXagent. To reduce the scope and complexity of the report generation task, we focus on generating the findings section only for each of the different CXR agents as this requires no clinical history. As described in section 
\ref{sec:report_generation_plan}, we define a set of metrics for evaluation based on Google's recent work \cite{yang2024Med-Gemini} and discussions with our clinical partners. We collect Rouge-L metrics to illustrate the limitations of NLP-based metrics. In addition to the metrics defined in section \ref{sec:report_generation_plan}, we add a ranking metric given we are now comparing between 4 models, where the clinical partner will rank the generations factoring in the quality, brevity and patient management implications. Some key pointers on this ranking metric:
\begin{itemize}
    \item The rank facilitates easy comparison only, it does not mean any of the generations are good. The ranking should be considered in conjunction with the other metrics to get a more complete picture of the generated reports.
    \item Following on from the above point, we note that in the case of a \textbf{normal} scan all models were similar in generation quality hence the ranking was a less insightful metric.
    \item During preliminary trials aimed at refining our evaluation metrics and definitions, we encountered difficulties in clearly distinguishing between the lower-performing models as their responses were often clinically inaccurate and hallucinated. Rather than assigning scores directly equivalent to the numerical rankings, we implemented the following mapping system to convert ranks into scores: Ranks 4 or 3 map to score 1 to account for the similarity in poorer performing models; Rank 2 maps to score 2; and Rank 1 maps to score 3, with 3 being the highest and most favorable score.
\end{itemize}To maximise data collected in the time we have with each clinical expert (under 4 hours per expert), we build a custom evaluation system to speed up the data collection, see section \ref{sec:evaluation_data_collection_platform}.

\section{Pathology Detection \& Classification}
\label{sec:evaluation_pathology_detection_classification}
In table \ref{tab:chexpert_agent_pathology_exact_match} and \ref{tab:vindr_agent_pathology_exact_match}, one can see two different CheXagent models, a ViT probe and three different agents. To clarify:
\begin{itemize}
    \item \textbf{ViT probe} is the concrete implementation of the pathology detector interface, it shows the drop in performance between the probe output and the overall output due to the LLM generation engine
    \item \textbf{CheXagent*} is the CheXagent model from figure \ref{fig:exact_match_CheXagent} with the highest exact match accuracy, namely using prompt "What pathologies are in the image?" with temperature 0.5
    \item \textbf{CheXagent (Temp = 1)} is a CheXagent model running on the same prompt as the agents, namely "Just list the findings on the chest x-ray, nothing else. If there are no findings, just say that" with temperature 0.5.
    \item \textbf{\{CheXagent, Llama 3, Gemini\} Agents} are the CXR-agents running with the ViT probe and the different generation engines listed in section \ref{sec:generation_engines}.
\end{itemize}

\begin{longtable}{>{\columncolor{gray!10}}>{\centering\arraybackslash}m{4.5cm} >{\centering\arraybackslash}m{2cm} >{\centering\arraybackslash}m{2cm} >{\centering\arraybackslash}m{2cm} >{\centering\arraybackslash}m{2cm}}
\rowcolor{gray!20} \multicolumn{5}{c}{\textbf{Exact Match Accuracies (\%)}} \\ 
\rowcolor{gray!20} \textbf{Model} & \textbf{Overall} & \textbf{No Finding} & \textbf{One Pathology} & \textbf{Multiple Pathology} \\
\hline
\endfirsthead

\rowcolor{gray!20} \multicolumn{5}{c}{\textbf{Exact Match Accuracies (\%)}} \\ 
\rowcolor{gray!20} \textbf{Model} & \textbf{Overall} & \textbf{No Finding} & \textbf{One Pathology} & \textbf{Multiple Pathology} \\
\hline
\endhead
ViT probe & 26.0 & 78.0 & 31.0 & 3.0 \\ 
CheXagent* & 15.7 & 55.0 & 3.3 & 2.0 \\ \hline
CheXagent (Temp = 1) & \textbf{25.0} & \textbf{100.0} & 0.0 & 0.0 \\ 
CheXagent Agent & \textbf{25.0} & \textbf{100.0} & 0.0 & 0.0 \\ 
Llama 3 Agent & 21.0 & 70.0 & \textbf{13.0} & \textbf{3.0} \\ 
Gemini Agent & 21.0 & 70.0 & \textbf{13.0}& \textbf{3.0} \\ 
\caption{Comparison of agents on \textbf{CheXpert} using prompt: "Just list the findings on the chest x-ray, nothing else. If there are no findings, just say that."; * = we compare against the best performing model in figure \ref{fig:exact_match_CheXagent}. Please refer to section \ref{accuracy-definitions} to see accuracy definitions.} 
\label{tab:chexpert_agent_pathology_exact_match} 
\end{longtable}

\begin{longtable}{>{\columncolor{gray!10}}>{\centering\arraybackslash}m{3cm} >{\centering\arraybackslash}m{2cm} >{\centering\arraybackslash}m{2cm} >{\centering\arraybackslash}m{2cm} >{\centering\arraybackslash}m{2cm} >{\centering\arraybackslash}m{2.5cm} >{\centering\arraybackslash}m{2.5cm}}
\rowcolor{gray!20} \multicolumn{5}{c}{\textbf{Exact Match Accuracies (\%)}} \\ 
\rowcolor{gray!20} 
\textbf{Model} & \textbf{Overall} & \textbf{No Finding} & \textbf{One Pathology} & \textbf{Multiple Pathology} & \textbf{Single Match Pathology Accuracy} \\ \hline
\endfirsthead

\rowcolor{gray!20} 
\textbf{Model} & \textbf{Overall} & \textbf{No Finding} & \textbf{One Pathology} & \textbf{Multiple Pathology} & \textbf{Single Match Pathology Accuracy} \\ \hline
\endhead
ViT probe & 71.0 & 98.0 & 29.0 & 5.0 & 66.0 \\
CheXagent* & 25.3 & 36.7 & 2.7 & 0.0 & 10.3 \\ \hline
CheXagent (Temp = 1) & \textbf{68.0}& \textbf{100.0} & 2.0 & \textbf{1.0 }& 3.0 \\ 
CheXagent agent & \textbf{68.0} & 99.0 & 0.0 & 0.0 & 0.0 \\ 
Llama 3 agent & 61.0 & 86.0 & \textbf{21.0} & 0.0 & \textbf{21.0} \\ 
Gemini agent & 62.0 & 87.0 & \textbf{21.0} & 0.0 & 20.0 \\ 

\caption{Comparison of agents on \textbf{VinDr} with exact match accuracies using prompt: "Just list the findings on the chest x-ray, nothing else. If there are no findings, just say that."; * = we compare against the best performing model in figure \ref{fig:exact_match_CheXagent}. Please refer to section \ref{accuracy-definitions} to see accuracy definitions.} 
\label{tab:vindr_agent_pathology_exact_match}
\end{longtable}

Tables \ref{tab:chexpert_agent_pathology_exact_match} and \ref{tab:vindr_agent_pathology_exact_match}, show many models across many metrics. Here we breakdown our key interpretations:

\begin{enumerate}
    \item \textbf{Low CheXagent accuracies on abnormal scans (i.e. scans with pathologies)}: Whilst CheXagent (Temp = 1) and CheXagent Agent show the highest overall performance, they have zero accuracy on the one and multiple pathology splits on CheXpert. For VinDr, CheXagent (Temp = 1) has 3\% single match pathology accuracy showing that of all pathologies present across all scans, it has only identified 3\% correctly relative to 66\% of the ViT probe and 20+\% of the Llama 3/Gemini agents. The bottom line:
    \begin{enumerate}
        \item Overall accuracy alone is insufficient to evaluate these models, one should look at the splits by normal and abnormal scans. We find single match pathology accuracy to be an insightful metric when there is little to separate between the accuracies of abnormal cases
        \item CheXagent models and CheXagent agent underperform on abnormal cases relative to the Llama 3 and Gemini agents
    \end{enumerate}
    \item \textbf{All tested LLM generation engines cause accuracy to decrease relative to their inputs from the ViT probe}: We see a reduction in performance from the ViT probe to the agents due to their LLM generation engines (ignoring CheXagent agent as this seems to overpredict no finding). However, there is no clear trend of greater reduction on normal vs abnormal cases, since on CheXpert we see the greatest reduction in the cases with one pathology but on VinDr we see the greatest reduction on the normal scans.
    \item \textbf{Negligible difference in performance between best performing Llama 3 and Gemini agents}: Llama3 and Gemini show the strongest performance on cases with pathologies (aside from CheXagent (Temp = 1)'s multiple pathology accuracy). However, we see that the underlying generation engine has very little impact (at most 1\% difference) on pathology detection across normal, single pathology and multiple pathology cases. Interestingly, the fine-tuned Mistral LLM does seem to perform much worse on abnormal cases, however this model is not representative of general medical fine-tuning as its been fine-tuned over MIMIC CXR reports in addition to its base medical fine-tuning.
\end{enumerate}

\section{Pathology Localisation}
\label{sec:evaluation_pathology_localisation}

For pathology localisation, we only focussed on the left vs right cases in the CheXbench image-text reasoning test set, however when inspecting the observations in these scans we noticed that not all observations would be detected by the pathology detector as they were not a label of either CheXpert or VinDr's pathologies, for example the pathology detector was unable to detect an elevated diaphragm, indwelling catheter, thoracic vertebrae scoliosis to name a few. Since our pathology localisation workflow requires the pathology to be detected by the pathology detector first before it is passed to the phrase grounding tool to be localised, we cannot localise all the left vs right test cases in CheXbench. This is a limitation of our architecture, which is designed for findings generation rather than question answering. In the future, we would adopt a more agent-like workflow by telling the LLM what tools it has at its disposal and asking it to use them to answer user questions, which would work better on this test benchmark assuming similar performance to what is seen in table \ref{tab:localisation_accuracy}. \\ 

\begin{figure}
    \centering
    \includegraphics[width=1\linewidth]{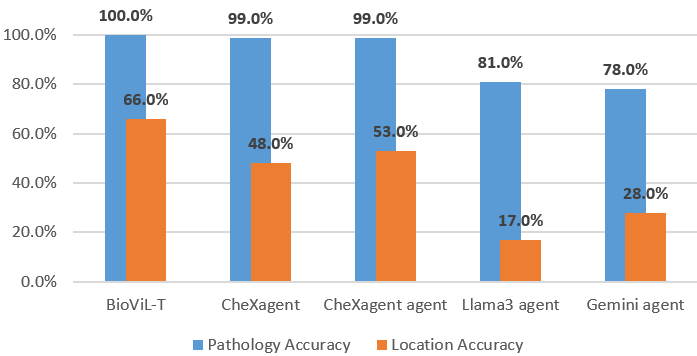}
    \caption{Agent pathology localisation results on CheXbench's left vs right cases over \{opacity, atelectasis, effusion\} pathologies only}
    \label{fig:agent_pathology_localisation}
\end{figure}

From figure \ref{fig:agent_pathology_localisation}, it is clear that for the LLama3 and the Gemini Agent models the pathology being localised is only being detected in 4 out of 5 cases. This is surprising given we have already cut down the test set to contain only three pathologies ("opacity", "atelectasis", "effusion"). When comparing this with BioViL-T on the same set,  we see that the pathology detector is bottlenecking the performance. This is likely because it only returns the label (i.e. atelectasis) to which we prepend a location (i.e. right atelectasis) before passing to our BioVil-T phrase grounding pipeline. However, in the CheXbench evaluation dataset, we have the complete option (i.e. right base pulmonary atelectasis) which contains more information that potentially helps increase the phrase grounding confidence and as a result, the localisation accuracy. We are assuming the CheXagent agent is hallucinating as the information it receives in its prompt is the same as the other two agents so it is unlikely to perform this much better and since this is a binary selection task, a score of close to 50\% suggests no better performance than random guessing.\\

Ultimately, our current workflow for pathology localisation is sub-optimal. We note that BioViL-T is a strong phrase grounding tool, however to get its best performance we need to give it more complete phrases and the current pathology detection probes do not provide this data. A good next step would be to train probes explicitly for pathology localisation   (i.e. have a left and right output neuron per pathology). We did not do this originally as BioViL-T showed promising pathology localisation results relative to CheXagent and even though the LLM has been shown to bottleneck performance, the questions were close-ended with two options hence we did not expect performance to be bottlenecked as significantly as open-ended questions/tasks. In the future, it would be good to have these localisation probes and compare their performance against BioViL-T. 

\section{Report Generation}
\label{sec: report-generation-overall-evaluation}

We first provide an overview of the platform built to aid data collection with our clinical experts, before looking at the key data from the evaluations. We split this review of key data by normal and abnormal scans as all 4 models were very good at distinguishing when a scan was normal from abnormal, however, the quality of generated reports varied significantly for abnormal scans relative to normal scans.

\subsection{Evaluation Data Collection Platform for Clinical Experts}
\label{sec:evaluation_data_collection_platform}
\begin{figure} [h!]
    \centering
    \includegraphics[width=0.6\linewidth]{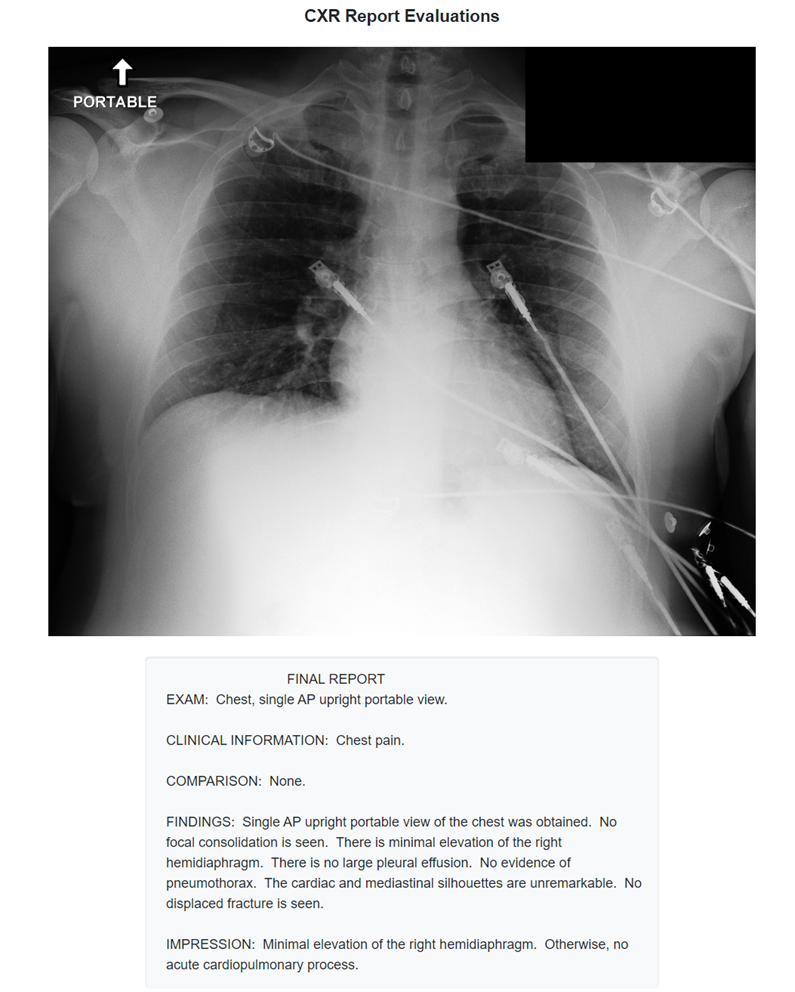}

    \adjustbox{lap=-0.15\linewidth}{\includegraphics[width=1.3\linewidth]{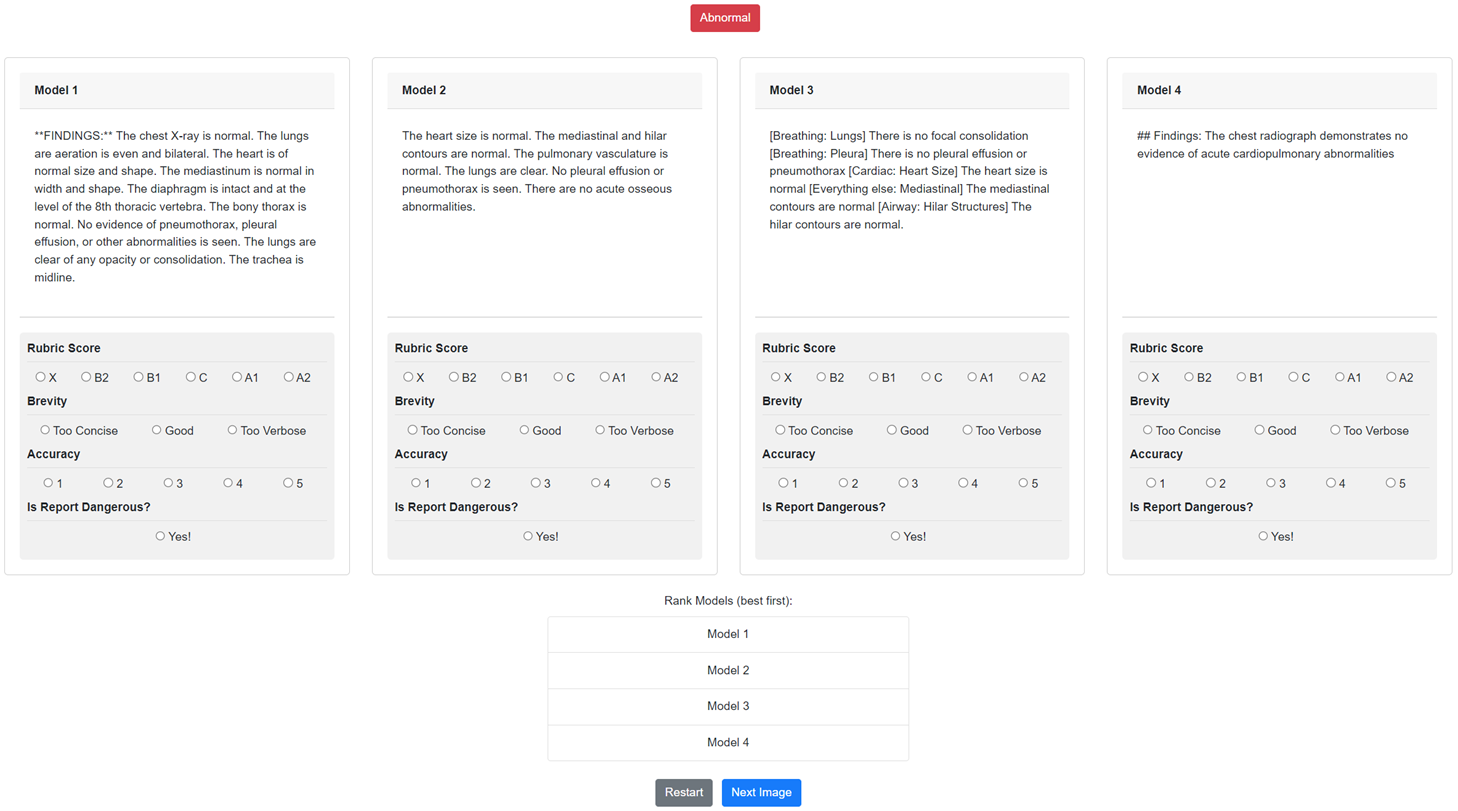}}
    \caption{A screenshot of our web application to collect data from clinical partners}
    \label{fig:data-collection-platform}
\end{figure}

We were fortunate to work with Dr Dominic Marshall (DM), an ST6 respiratory specialist registrar and an Honorary Clinical Research Fellow at the University of Oxford, as well as Dr Yueqi Ge (YG), an ST4 respiratory specialist registrar. Given the time constraints of our clinical partners we built a bespoke data collection platform (see figure \ref{fig:data-collection-platform}) to collect the rankings and all other metrics outlined in section \ref{sec:report_generation_plan} for the four different models (including CheXagent). \\

When building this web-based data collection platform our objectives were:
\begin{itemize}
    \item \textbf{Simplicity}: clinical partners did not have time to learn to use a complicated data collection system
    \item \textbf{Bias reduction}: we ensured all models were anonymised and randomly shuffled between their labels (i.e. model 1 did not always map to the same underlying generation engine)
    \item \textbf{Completeness}: we wanted a single platform to collect all metrics we were interested in with the abnormal button allowing us to split the data between normal and abnormal findings offering valuable post-collection insights
\end{itemize}

We built this web application with Python's Flask framework using a Bootstrap toolkit for responsive design. A key issue that came up was the generation time for each model's report (roughly 1 minute for all 4 models altogether), we overcame this by simply pre-collecting all the reports as our prompt was fixed once we had found the best one per model (see section \ref{sec:generation_engines}). The user prompt we selected was "What are the findings?".

\subsection{Normal scans}

\begin{longtable}{>{\columncolor{gray!10}}>{\centering\arraybackslash}m{3.5cm} >{\centering\arraybackslash}m{2.55cm} >{\centering\arraybackslash}m{2cm} >{\centering\arraybackslash}m{2.05cm} >{\centering\arraybackslash}m{2.55cm}}
\rowcolor{gray!20}
\textbf{Model} & \textbf{Reference report comparison $\uparrow$} & \textbf{Brevity*} & \textbf{Accuracy  $\uparrow$} & \textbf{Dangerous report count $\downarrow$} \\ \hline

\endfirsthead

\rowcolor{gray!20}
\textbf{Model} & \textbf{Reference report comparison $\uparrow$} & \textbf{Brevity*} & \textbf{Accuracy  $\uparrow$} & \textbf{Dangerous report count $\downarrow$} \\ \hline

\endhead

CheXagent & \textbf{-0.10} & \textbf{-0.05} & 4.50 & 0 \\
CheXagent agent & -0.11 & 0.2 & \textbf{4.56} & 0 \\
Llama3 agent & -0.50 & -0.30 & 3.40 & 0 \\
Gemini agent & -0.65 & -0.55 & 3.40 & 0 \\

\caption{Evaluation metrics of CheXagent and the different generation engines for normal MIMIC-II scans (all metrics shown are averages from both registrars); * for brevity -1 is too concise, 0 is good and 1 is too verbose}
\label{tab:normal_reports_eval}
\end{longtable}

From sitting through the evaluation process with YG and DM, and based on the collected data, we immediately note that for normal scans there is very little to separate the models aside from brevity. Given the variation in reports of normal scans, we do not heavily penalise concise reports, as this is how some radiologists would report. Interestingly, the less experienced respiratory registrar (YG) preferred the longer reports for the normal scans, as Llama 3 and Gemini had average scores of 3 for accuracy relative to the more experienced registrar's 3.8, accuracy being the biggest \% difference on any metric between the two registrars for normal scans. This shows that brevity can impact accuracy as more interpretation is required from the doctor reading the report. However, the clinical implications are the same with no impact on patient management based on all averaged accuracy scores being over 3 with reference report comparisons being B1 or better (recall B1 implies correct patient management but fewer relevant clinical findings than reference report).

\subsection{Abnormal Scans}

\begin{longtable}{>{\columncolor{gray!10}}>{\centering\arraybackslash}m{3.5cm} >{\centering\arraybackslash}m{3.5cm} >{\centering\arraybackslash}m{3.5cm} >{\centering\arraybackslash}m{3.25cm}}
\rowcolor{gray!20}
\textbf{Model} & \textbf{MIMIC-II reference report comparison $\uparrow$} & \textbf{MIMIC-II temporal hallucinations $\downarrow$} & \textbf{CheXpert temporal hallucinations $\downarrow$} \\ \hline
\endfirsthead
\rowcolor{gray!20}
\textbf{Model} & \textbf{MIMIC-II reference report comparison} & \textbf{MIMIC-II temporal hallucinations} & \textbf{CheXpert temporal hallucinations} \\ \hline
\endhead
CheXagent & -1.24 & 11 (44\%) & 15 (79\%) \\
CheXagent agent & -1.34 & 0 & 0 \\ 
Llama3 agent & -1.02 & 0 & 0 \\ 
Gemini agent & -0.98 & 0 & 0 \\ 
\caption{Reference report and temporal hallucination comparison between CheXagent and the the different generation engines}
\label{tab:temporal_hallucinations}
\end{longtable}

\begin{figure}
    \centering
    \includegraphics[width=1\linewidth]{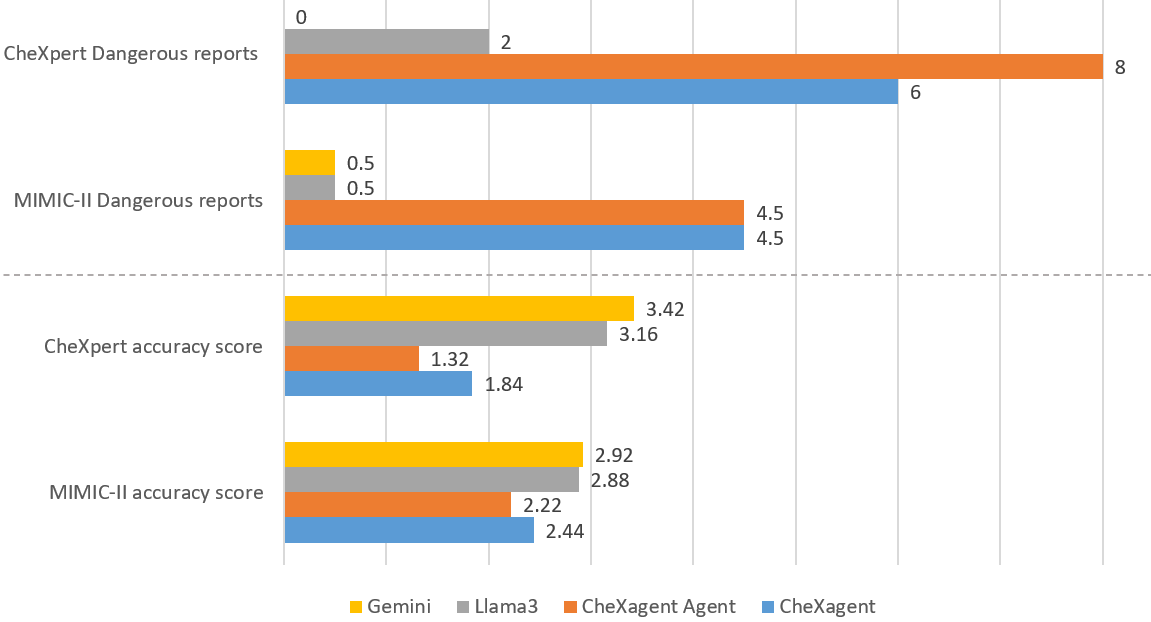}
    \caption{Comparison of CheXagent and the different generation engine's accuracy scores and dangerous report counts across MIMIC-II and CheXpert}
    \label{fig:abnormal_reports_accuracy_and_danger}
\end{figure}

\begin{longtable}{>{\columncolor{gray!10}}>{\centering\arraybackslash}m{3.5cm} >{\centering\arraybackslash}m{3.5cm} >{\centering\arraybackslash}m{3.5cm} >{\centering\arraybackslash}m{3.25cm}}
\rowcolor{gray!20}
\textbf{Model} & \textbf{AI superior/similar to original report (Normal)} & \textbf{AI superior/similar to original report (Abnormal)} & \textbf{Difference between normal and abnormal} \\ \hline

\endfirsthead

\rowcolor{gray!20}
\textbf{Model} & \textbf{AI superior/similar to original report (Normal)} & \textbf{AI superior/similar to original report (Abnormal)} & \textbf{Delta} \\ \hline

\endhead

Med-Gemini & 57\% & 43\% & -14 \\ \hline
CheXagent & 90\% & 24\% & -66 \\
CheXagent agent & 85\% & 16\% & -69 \\
Llama3 agent & 50\% & 20\% & -30\\
Gemini agent & 45\% & 20\% & -25 \\

\caption{Rubric evaluation compared to Med-Gemini \cite{yang2024Med-Gemini}, where any report scoring C(0), A1(1), A2(2) is considered similar to superior to original report, average of DM and YG}
\label{tab:med_gemini_comparison}
\end{longtable}

\begin{longtable}{>{\columncolor{gray!10}}>{\centering\arraybackslash}m{3.5cm} >{\centering\arraybackslash}m{3.5cm} >{\centering\arraybackslash}m{3.5cm} >{\centering\arraybackslash}m{3.25cm}}
\rowcolor{gray!20} \multicolumn{4}{c}{\textbf{Rouge-L (\%)}} \\ 
\rowcolor{gray!20}
\textbf{Model} & \textbf{All} & \textbf{Normal} & \textbf{Abnormal} \\ \hline
\endfirsthead
\rowcolor{gray!20} \multicolumn{4}{c}{\textbf{Rouge-L(\%)}} \\ 
\rowcolor{gray!20}
\textbf{Model} & \textbf{All} & \textbf{Normal} & \textbf{Abnormal} \\ \hline
\endhead

CheXagent & 10.7\% & 14.8\% & 9.0\% \\
CheXagent agent & 4.9\% & 7.6\% & 3.7\% \\
Llama3 agent & 11.3\% & 9.8\% & 11.9\% \\
Gemini agent & 17.4\% & 20.7\% & 16.1\% \\
\caption{Rouge-L (Findings + Impression) scores comparison}
\label{tab:Rouge-L}
\end{longtable}

On the other hand, there is significant variation between the different models for abnormal scans. The bottom line across reports generated by all models is that they are worse than radiologist reports based on table \ref{tab:temporal_hallucinations}, where we see the average score of all models being below 0 in terms of reference report comparisons recalling that -1 maps to B1 and 0 maps to C in the comparison rubric (see rubric: \ref{fig:google_report_comparison_rubric}). However, we note the Gemini agent and Llama3 agent are more likely to lead to correct patient management than CheXagent and CheXagent agent. \\

In general, based on figure \ref{fig:abnormal_reports_accuracy_and_danger} we see a clear trend across both MIMIC-II and CheXpert with the Gemini agent being the best performing model, in terms of fewest dangerous reports and highest accuracy; we note that Llama3 is close behind. However, there is a significant difference between these two models and CheXagent/CheXagent agent, particularly on the CheXpert dataset. This suggests that the CheXagent LLM is overfitting to MIMIC-CXR given the drop-off in performance, which is likely because it was fine-tuned over MIMIC-CXR. The Rouge-L metrics in table \ref{tab:Rouge-L} highlight this since CheXagent and Llama3 appear very similar in their performance however, we know from table \ref{tab:temporal_hallucinations} and figure \ref{fig:abnormal_reports_accuracy_and_danger} that CheXagent hallucinates far more frequently and produces many more dangerous reports. This helps illustrate our earlier point on Rouge-L and NLP metrics in general being insufficient for report evaluation.\\

Our agent-based workflow generalises far better, since its accuracy scores are higher on CheXpert. We note Gemini outperforms Llama3, suggesting generation engines with larger LLMs (i.e. more parameters) are better for these medical agents although we think this would have diminishing returns since the disparity in performance is not significant. Unfortunately, we are unable to properly assess the impacts of fine-tuning on the LLM, since CheXagent agent's Mistral LLM scores very poorly, however, this is likely due to it being fine-tuned on MIMIC-CXR reports specifically. An interesting future step would be to use a fine-tuned medical LLM but one that has not overfit to a specific chest x-ray dataset.\\

From table \ref{tab:temporal_hallucinations}, it is evident that CheXagent hallucinates significantly on abnormal scans; in particular, many generated reports contain references to non-existent priors or changes in observations, when we only provide a single test scan. This is likely due to the limited diversity in reports used for fine-tuning (only the MIMIC-CXR set) resulting in overfitting, since nearly 80\% of the MIMIC-CXR  reports refer to prior data. This correlates highly with the number of CheXpert reports with temporal hallucinations, which is also almost 80\%. In practise, it is likely that the appearance of certain pathologies in a scan and the appearance of certain terminology in the report has resulted in the model learning incorrect correlations between scan and report. This is further motivation for causal analysis in medicine. \\

Alternately, a way to tackle these hallucinations and overfitting would be to include more training data from a variety of clinical settings, this is how Med-Gemini was fine-tuned as they had access to a private dataset of paired reports and scans from an Indian hospital, in addition to MIMIC-CXR. When comparing the rubric scores between Med-Gemini, CheXagent and the various agents in table \ref{tab:med_gemini_comparison}, its clear that this additional training data reduces overfitting due to the reduction in performance difference between normal and abnormal scans, as well as the far higher performance on abnormal scans. \\

\section{Key Takeaways}
We conclude our evaluation of the various CXR agents by presenting our key takeaways which may help guide future work.

\begin{itemize}
    \item \textbf{Metrics:} When evaluating models on the various downstream tasks it was insightful to look at results split between normal and abnormal scans, since results for all models on abnormal scans were almost always worse. Hence when comparing models it was often better to look at their performance on abnormal scans, particularly in the case of generated reports, to assess the impacts of accuracy and hallucination on patient management. For report evaluation, Rouge-L metrics were insufficient as they showed Llama3 and CheXagent to be similar even though CheXagent had far lower accuracy and produced many more acutely dangerous reports. For pathology detection, we find single match pathology accuracy useful when there is little to separate between the exact match accuracies of abnormal cases as this shows when many pathologies are correctly identified in isolation (refer to section \ref{accuracy-definitions} for accuracy definitions). Refer to tables \ref{tab:chexpert_agent_pathology_exact_match}, \ref{tab:vindr_agent_pathology_exact_match}, \ref{tab:med_gemini_comparison} and \ref{tab:Rouge-L}.
    
    \item \textbf{For pathology detection:} the LLM bottlenecks probe performance regardless of which LLM is used; however, this reduction in performance is lower for agent-based workflows, excluding  CheXagent agent's overfitted LLM. Ultimately if the downstream task is pathology detection and classification only, it is better to use encoders trained at a foundational scale (i.e. across many datasets with different pathologies from different countries). Otherwise, an intelligent (based on parameter count), general LLM performs best when summarising information from a probe. However, in the future, we would like to evaluate against a medical fine-tuned LLM that has not overfit to MIMIC-CXR. Refer to section \ref{sec:evaluation_pathology_detection_classification}, tables \ref{tab:chexpert_agent_pathology_exact_match} and \ref{tab:vindr_agent_pathology_exact_match}.

    \item Our current setup of the CXR agent limits \textbf{pathology localisation} performance as we only ground pathologies generated by the pathology detector meaning we cannot localise pathologies that have not been detected. This is due to our architecture being optimised for findings generation rather than question answering. The pathology localisation evaluations in the CheXbench benchmark are question-answering based. A future step would be to adapt our design using agent-specific libraries (i.e. Langchain) to allow the LLM to use the pathology detector and localiser as functions which can be called based on the user prompt rather than functions which are called in a specific order regardless of the user prompt. Refer to section \ref{sec:evaluation_pathology_localisation}.

    \item \textbf{Report generation:} For abnormal scans, the majority (over 55\%) of AI-generated reports are worse than radiologists due to hallucinations, the lack of granularity when describing pathologies and limitations on support device related findings. However, our best CXR agents, namely Gemini agent and Llama3 agent, are better than CheXagent at report generation as they are more accurate and generate very few acutely dangerous reports. Their standout attribute is the generalisation ability with generated reports being better on CheXpert than MIMIC-II, unlike CheXagent which demonstrates clear overfitting with its drop in performance on CheXpert. This overfitting is further demonstrated by CheXagent agent's significantly lower performance compared to the Llama3 and Gemini agents, indicating that the underlying Mistral LLM has been excessively fine-tuned on the MIMIC-CXR reports. We think this overfitting can be tackled by training over more datasets of a similar scale to MIMIC-CXR but from different medical settings. Refer to section \ref{sec: report-generation-overall-evaluation}, table \ref{tab:med_gemini_comparison} and figure \ref{fig:abnormal_reports_accuracy_and_danger}.

\end{itemize}

\chapter{Conclusion and Future Work}
\section{Conclusion}
We have thoroughly evaluated the performance of publicly available, state-of-the-art models like CheXagent and BioViL-T on various tasks related to chest X-ray analysis in chapter \ref{chap: sota-investigation}. We find that vision encoders, extracted from foundational, multimodal models using linear probing, outperform current state-of-the-art models like Torch X-ray vision and CheXagent for pathology detection and classification tasks. These vision encoders demonstrate strong generalization abilities across different datasets and domain shifts. We observe that the overall performance of CheXagent is bottlenecked by its language model component, which frequently hallucinates and shows signs of overfitting due to it mostly being trained on the MIMIC-CXR dataset. Most importantly, we find that confident language when hallucinating significantly affects clinical interpretation, due to time spent carefully validating these confident yet incorrect statements. Hence we aim to incorporate uncertainty-aware language to improve clinical interpretation. \\

To address these limitations in chapter \ref{chap: cxr-agent}, we propose an agent-based vision-language workflow that leverages the probed vision encoder and phrase grounding tools to generate prompts with pathologies and their locations. These prompts are used by a language model to generate the findings sections of radiology reports with language factoring in uncertainty related to observations. By experimenting with various language models as generation engines, we find that our best agents outperform CheXagent, the current publicly available state-of-the-art model for findings generation, while also demonstrating far better generalization across different datasets. We acknowledge limitations in our pathology localisation workflows and realise this bounds our interpretability, though we suggest solutions including function calling agents and training of localisation probes. \\

Our investigation highlights the importance of evaluating models on both normal and abnormal scans, as well as placing heavy emphasis on clinical evaluations due to the risks of evaluating solely on NLP-based metrics. Finally, we emphasize the need for more diverse training data and potential report augmentation techniques to mitigate hallucinations and overfitting issues observed in existing models as this will help improve end-to-end LMM training in the long run. In conclusion, foundational vision encoders and improved LLM integration are crucial for enhancing pathology detection, classification, and report generation in medical imaging.

\section{Future work}
We break down future work by complexity and time required.

Simpler work:
\begin{itemize}
   \item \textbf{Probes for pathology localisation}: Our early investigations of probes for pathology localisation did not show huge promise though we did not experiment thoroughly. Given our framework to collect embeddings and the VinDr dataset we have with pathologies locations labelled, an interesting next step could be to train probes against this data and evaluate thoroughly.
   \item \textbf{Implement function calling in the agent} as this would allow for better phrase grounding and more versatile user Q\&A since the LLM would pick the tool based on the prompt
   \item \textbf{Evaluate the relationship between linear probe pathology activation and severity}: Using CheXbench one can try to determine whether there is any statistically significant correlation between the severity of a pathology and the confidence returned by a linear probe for that pathology. This is important since severity forms a key part of the generated radiology report.
   \item \textbf{Use a general medical fine-tuned LLM for the generation engine}, one that has not been overfit on CXR reports but a general medical corpus, such that it retains good general reasoning, and evaluate its performance across our benchmarked tasks. This was meant to be the point of CheXagent's Mistral LLM however this was suboptimal due to its overfitting.
\end{itemize}

More complex extensions:
\begin{itemize}
   \item \textbf{Train more probes for the various common Q\&A tasks}. Training more probes will allow for CXR agents to answer a greater variety of questions, for example: PA vs AP differentiation, placement of support devices and other benchmarked tasks. Ultimately, a foundational LMM should be able to achieve these tasks anyhow, given it is trained carefully on diverse data; currently, however, to maximise interpretability probes may be a better approach.
   \item \textbf{Fine-tune CheXagent on more large-scale datasets} (i.e. 100,000+ scans and reports ideally from different datasets than MIMIC-CXR) to reduce the overfitting and hallucinations. 
   \item \textbf{Longitudinal scans} form the most common types of scans in ICU settings,  currently CheXagent supports multiple input scans but we have not thoroughly evaluated its performance on this. One could explore an extension of uncertainty-aware radiology reporting agents to these scans.
\end{itemize}


\bibliographystyle{unsrtnat}
\bibliography{background}

\end{document}